\documentclass[a4paper,11pt]{article}
\usepackage{jcappub}

\usepackage{booktabs}
\usepackage{amsmath}
\usepackage{amssymb}
\usepackage{color}
\usepackage[dvipsnames]{xcolor}
\usepackage{todonotes}
\usepackage[normalem]{ulem}

\newcommand{\Neff}
{\ensuremath{N_{\rm eff}}}

\newcommand{\fortepiano}{\texttt{FortEPiaNO}}

\arxivnumber{2602.23915} 

\title{Impact of non-standard neutrino-electron interactions on Big Bang Nucleosynthesis}

\author[a]{Julien Froustey,}
\emailAdd{julien.froustey@ific.uv.es}
\author[a,b]{Stefano Gariazzo,}
\emailAdd{gariazzo@ific.uv.es}
\author[a]{Jaume Moncho,}
\emailAdd{jaumon@ific.uv.es}
\author[a]{Sergio Pastor}
\emailAdd{pastor@ific.uv.es}
\author[c,d]{and Ofelia Pisanti}
\emailAdd{pisanti@na.infn.it}

\affiliation[a]{Instituto de F{\'\i}sica Corpuscular (IFIC), CSIC-Universitat de Val{\`e}ncia, Parc Científic UV, C/ Catedrático José Beltrán 2, E-46980 Paterna (València), Spain}
\affiliation[b]{University of Turin, Physics department and INFN, Sezione di Torino, Via P. Giuria 1, I--10125 Torino, Italy}
\affiliation[c]{Dipartimento di Fisica “Ettore Pancini”, Universit\`a degli studi di Napoli “Federico II”, Complesso Univ. di Monte S. Angelo, I-80126 Napoli, Italy}
\affiliation[d]{INFN - Sezione di Napoli, Complesso Univ. di Monte S. Angelo, I-80126 Napoli, Italy}

\abstract{Neutrino non-standard interactions (NSI) with electrons, predicted in many extended theoretical models of particle physics, are known to alter the picture of neutrino decoupling from the cosmic plasma. We update previous analyses of neutrino decoupling in the presence of NSI with electrons, extending the parameter space and assessing their impact on the primordial light-element abundances produced during Big Bang Nucleosynthesis (BBN). We show that tracking the NSI-induced modifications of the neutrino distributions is necessary to accurately predict the helium-4 abundance, which cannot be derived from a pure $\Neff$-driven change. BBN therefore provides additional observables that are complementary to $\Neff$. We then derive cosmological constraints on non-universal and flavour-changing NSI parameters, using different dataset combinations, and compare them with bounds from terrestrial experiments. We find that the cosmological limits are significantly less stringent than the terrestrial ones, but they provide an independent probe of neutrino interactions in the early universe.}

\begin{document}

\maketitle

\flushbottom

\section{Introduction}
Current data from solar, atmospheric, reactor and accelerator neutrino experiments are well described by the existence of flavour oscillations
in the framework of three-neutrino mixing (see e.g.\ \cite{deSalas:2020pgw}).
This experimental evidence for non-zero neutrino masses and mixing calls for new physics beyond the Standard Model (SM) of fundamental particles and forces.
The presence of non-standard interactions (NSI) of neutrinos with other fermions and among themselves is predicted by a variety of extended theoretical models.
Such new interactions can leave an imprint on different observations and experiments,
including the pattern of neutrino oscillations in matter, neutrino scattering at production and detection, and neutrino interaction measurements.
By analysing a combination of all available data, there is currently no evidence in favour of NSI, and bounds have been derived, as reviewed e.g.\ in \cite{Farzan:2017xzy}, but their
existence would have implications in cosmological and astrophysical scenarios.

In the early universe, neutrinos are kept in equilibrium by weak interactions, which become ineffective at a temperature of ${\cal O}$(MeV).
At this point, neutrinos decouple and constitute the cosmological neutrino background.
The neutrino decoupling process is very well understood, see e.g.~\cite{Akita:2020szl,Froustey:2020mcq,Bennett:2020zkv},
where the effective number of relativistic species has been calculated to be $\Neff=3.044$.
This quantity is defined in order to quantify the amount of energy density ($\rho_{\rm rad}$)
carried by relativistic, neutrino-like relics 
with respect to the photon one ($\rho_\gamma$) \cite{Lesgourgues:2018ncw}:
\begin{equation}
\rho_{\rm rad} = \rho_\gamma \left[1+\frac{7}{8}\left(\frac{4}{11}\right)^{4/3}\,\Neff\right]\, .
\label{eq:neff}
\end{equation}
Measurements of the cosmic microwave background (CMB) anisotropies, combined with other cosmological data, provide constraints on \Neff. The \emph{Planck} collaboration obtained $\Neff=2.99 \pm 0.17$ at 68 \% C.L.~\cite{Aghanim:2018eyx}, while the recent combination with the Atacama Cosmology Telescope (ACT) and South Pole Telescope (SPT) data gives the tighter constraint $\Neff = 2.82 \pm 0.12$ at 68 \% C.L.~\cite{SPT-3G:2025bzu}.\footnote{We note that an independent analysis of a similar dataset combination~\cite{Tristram:2025you} reports a larger central value $\Neff =  3.031$ with $\pm 0.130$ statistical uncertainty, fully consistent with the standard prediction.} Such values severely restrict the possible existence of additional relativistic particles. 

The presence of NSI could modify the decoupling process of cosmological neutrinos, as noted already in \cite{Berezhiani:2001rs,Davidson:2003ha}. In particular,
non-standard interactions with electrons, the only charged leptons that are still abundant at MeV cosmological temperatures, would modify the thermal contact of neutrinos with the electromagnetic plasma and change their momentum spectra, leading to a different value of $\Neff$. Earlier analyses of the effect of NSI on $\Neff$ include \cite{Mangano:2006ar,deSalas:2016ztq}, but these works only considered some particular choices of the NSI parameters. 
More recently, the effect of NSI on neutrino decoupling has been quantified in \cite{deSalas:2021aeh},
where a complete numerical treatment of NSI in this context has been developed.

In contrast to the study of neutrino decoupling, a full analysis of NSI effects on Big Bang Nucleosynthesis (BBN) has not been performed yet. A partial study, involving non-standard interactions of neutrinos with quarks, was published recently in \cite{Barenboim:2025okj}. In this case, the weak interaction processes that govern the equilibrium of protons and neutrons at the beginning of BBN, the so-called weak rates, are modified.
This implies that the predicted abundances of light elements, and in particular helium-4, are modified by the ratio of neutrons with respect to protons at the onset of nucleosynthesis.

By contrast, if one considers four-fermion NSI between neutrinos and electrons alone, there is no direct effect on the weak rates (at tree level), but the presence of NSI modifies the total energy density of radiation in the universe and leads to distortions in the momentum distribution function of electron neutrinos. Thus, the effect on the light element abundances is different but it can still be appreciable. This is the objective of the present paper, in which we also compare the results with current observations of the primordial abundances.

The rest of this paper is organised as follows. We introduce our NSI parameterisation in section~\ref{sec:NSI}. The effect of NSI on neutrino decoupling, and in particular on the observable \Neff, is discussed in section~\ref{sec:results_Neff}. The subsequent impact of the modified relic neutrino distributions on primordial nucleosynthesis is determined in section~\ref{sec:results_BBN}, where we also determine the resulting constraints on the NSI parameters. We conclude in section~\ref{sec:conclusion}. Throughout this paper, we use natural units in which $\hbar = c = k_\mathrm{B} = 1$.

\section{Neutrino non-standard interactions with electrons}
\label{sec:NSI}

Non-standard neutrino interactions arise in a broad class of extensions of the SM that generate neutrino masses and mixing~\cite{Farzan:2017xzy}. At energies well below the electroweak scale, such effects can be described in terms of effective four-fermion operators that modify the interactions between neutrinos and matter fields. 

For the range of cosmological temperatures during neutrino decoupling, the only abundant charged leptons in the primordial plasma are electrons and positrons. Consequently, we restrict our analysis to neutral-current NSI of neutrinos with electrons. Taking this into account, the effective Lagrangian density describing neutrino--electron interactions can be written as
\begin{equation}
    \mathcal{L} = \mathcal{L}_{\rm SM} + \mathcal{L}_{\mathrm{NC-NSI}e} \, ,
\end{equation}
where the SM contribution reads
\begin{equation}
        \mathcal{L}_{\rm SM} = -2\sqrt{2}\,G_F \Big[(\bar{\nu}_e \gamma^\mu P_L e)(\bar{e} \gamma_\mu P_L \nu_e) + g_X\, (\bar{\nu}_\alpha \gamma^\mu P_L \nu_\alpha) (\bar{e} \gamma_\mu P_X e) \Big] \, .
\end{equation}
\noindent Here, \(G_F\) denotes the Fermi constant and $X = L,R$ is the chirality, so that \( P_{R,L} = (1 \pm \gamma_5)/2 \) are the usual right- or left-handed chirality projection operators, and \( g_L = \sin^2{\theta_W}-1/2, g_R = \sin^2{\theta_W} \) are the SM neutral-current couplings, with $\theta_W$ being the weak mixing angle. The NSI contribution is described by
\begin{equation}
\label{eq:L_NSI}
    \mathcal{L}_{\mathrm{NC-NSI}e} = -2\sqrt{2}\,G_F \thinspace \varepsilon_{\alpha\beta}^X \left(\overline{\nu}_{\alpha}\gamma^{\mu}P_L\nu_{\beta}\right) \left(\overline{e}\gamma_{\mu}P_X e\right)\thinspace,
\end{equation}
where the dimensionless parameters \(\varepsilon_{\alpha\beta}^{X}\) quantify the strength of the non-standard interactions between neutrinos of flavours \(\alpha\) and \(\beta\) and electrons. The sums over flavours $\alpha,\beta$ and chiralities $X$ are implicit in both equations.

We refer to NSI as \textit{flavour-changing} when \( \varepsilon_{\alpha\beta} \neq 0 \) for \( \alpha \neq \beta \), since they correspond to lepton flavour non-conserving interactions. On the other hand, if \( \varepsilon_{\alpha\alpha} - \varepsilon_{\beta\beta} \neq 0 \), we refer to them as \textit{non-universal} NSI, as they break lepton flavour universality.

Bounds on neutrino NSI with electrons can be derived from a combination of data from neutrino oscillation experiments, neutrino--electron scattering measurements, and precision constraints from LEP. Oscillation and scattering experiments constrain different combinations of NSI parameters through their impact on neutrino propagation and detection processes, respectively. 

In Table~\ref{tab:NSIbounds} we present a compilation of  bounds on neutrino non-standard interactions with electrons, obtained under the assumption that only one NSI parameter is non-zero at a time.
If correlations between different parameters are taken into account, the allowed parameter space becomes larger, because of degeneracies and cancellations that may appear among different NSI coefficients.
In order to obtain more robust bounds, one possibility is to vary several NSI parameters at the same time, and then to marginalise over some of them in order to present one-dimensional constraints.
We note that more recent analyses of NSI constraints have appeared in the literature, based on global fits that adopt a marginalisation procedure and employ alternative parameterisations of the interaction structure (see e.g.~\cite{Coloma:2023ixt}). 
The resulting bounds are therefore not directly comparable to those reported here. 
The limits summarised in our work are chosen for consistency with the one-parameter scans performed in our cosmological analysis and to facilitate a transparent comparison with previous studies.

\begin{table}[t!]
\centering
\caption{Current bounds on non-universal and flavour-changing NSI with electrons at 90\% C.L. for 1 degree of freedom. Adapted from \cite{Farzan:2017xzy, Martinez-Mirave:2023fyb}.}
\label{tab:NSIbounds}
\renewcommand{\arraystretch}{1.3}
\begin{tabular}{l@{\hskip 1cm}l}
\toprule
Parameter and 90\% C.L. range & Origin \\
\midrule
$-0.021 < \varepsilon_{ee}^L < 0.052$ & Neutrino oscillations \cite{Bolanos:2008km} \\
$-0.07 < \varepsilon_{ee}^R < 0.08$ & Scattering data \cite{Deniz:2010mp} \\
$-0.23 < \varepsilon_{ee}^R < 0.07$ & Neutrino oscillations \cite{Coloma:2022umy} \\
\addlinespace
$-0.03 < \varepsilon_{\mu\mu}^{L,R} < 0.03$ & Scattering and accelerator data \cite{Barranco:2007ej} \\
$-0.12 < \varepsilon_{\tau\tau}^L < 0.06$ & Neutrino oscillations \cite{Bolanos:2008km} \\
$-0.98 < \varepsilon_{\tau\tau}^R < 0.23$ & Neutrino oscillations \cite{Bolanos:2008km, Agarwalla:2012wf} \\
$-0.25 < \varepsilon_{\tau\tau}^R < 0.43$ & Scattering and accelerator data \cite{Bolanos:2008km} \\
\addlinespace\hline
\addlinespace
$-0.13 < \varepsilon_{e\mu}^{L,R} < 0.13$ & Scattering and accelerator data \cite{Barranco:2007ej} \\
$-0.33 < \varepsilon_{e\tau}^L < 0.33$ & Scattering and accelerator data \cite{Barranco:2007ej} \\
$\phantom{-}0.05 < \lvert\varepsilon_{e\tau}^R\rvert < 0.28$ & Scattering and accelerator data \cite{Barranco:2007ej} \\
$-0.19 < \varepsilon_{e\tau}^R < 0.19$ & Scattering data \cite{Deniz:2010mp} \\
$-0.10 < \varepsilon_{\mu\tau}^{L,R} < 0.10$ & Scattering and accelerator data \cite{Barranco:2007ej} \\
\bottomrule
\end{tabular}
\end{table}

\section{Impact of non-standard interactions on neutrino decoupling}
\label{sec:results_Neff}

\subsection{Quantum Kinetic Equations with NSI}

In order to consistently account for neutrino flavour oscillations, spectral distortions, and non-standard interactions during the decoupling process, we adopt the density matrix formalism, where the density matrix is defined as
\begin{align}
\varrho(t,p) = 
\begin{pmatrix} 
\varrho_{ee} & \varrho_{e\mu} & \varrho_{e\tau} \\ 
\varrho_{\mu e} & \varrho_{\mu\mu} & \varrho_{\mu\tau} \\ 
\varrho_{\tau e} & \varrho_{\tau\mu} & \varrho_{\tau\tau} 
\end{pmatrix} 
= 
\begin{pmatrix} 
f_{\nu_e} & \varrho_{e\mu} & \varrho_{e\tau} \\ 
\varrho_{\mu e} & f_{\nu_\mu} & \varrho_{\mu\tau} \\ 
\varrho_{\tau e} & \varrho_{\tau\mu} & f_{\nu_\tau} 
\end{pmatrix}\thinspace .
\end{align}
The diagonal elements are the momentum distribution functions of the different neutrino flavours, and the off-diagonal elements are non-zero in the presence of mixing and implement the coherence of the system.
Using the comoving variables (see e.g., \cite{Bennett:2020zkv,Gariazzo:2019gyi}) $x=m_e a$, $y=p a$, $z=T_\gamma a$, where $m_e$ is the electron mass, $p$ the neutrino momentum, $T_\gamma$ the photon temperature and $a$ the cosmological scale factor, the evolution of $\varrho$ is given by the quantum kinetic equation, written as
\begin{align}
    \frac{d\varrho(y)}{dx} = \sqrt{\frac{3m_{Pl}^2}{8\pi\rho}} \Bigg\{ 
    -i\frac{x^2}{m_e^3} \left[ 
        \frac{\mathbb{M}_\textup{F}}{2y} 
        - \frac{2\sqrt{2}G_F\thinspace y \thinspace m_e^6}{x^6} 
        \left( \frac{\mathbb{E}_\ell + \mathbb{P}_\ell}{m_W^2}
        +\frac{4}{3}\frac{\mathbb{E}_\nu}{m_Z^2} \right), 
        \varrho(y) \right] 
    + \frac{m_e^3}{x^4} \mathcal{I}(\varrho) 
    \Bigg\} \thinspace,
    \label{eq: matrixevol}
\end{align}
where $m_W$ and $m_Z$ are the masses of the gauge bosons $W^\pm$ and $Z$, $\rho$ is the total energy density of the universe, $\mathbb{E}_\ell$, $\mathbb{P}_\ell$ and $\mathbb{E}_\nu$ are matrices that represent the energy density and pressure of charged leptons and neutrinos, and $\mathcal{I}(\varrho)$ encodes all the scattering/annihilations of the system involving electrons and positrons and also neutrino-neutrino interactions (for details, see \cite{Gariazzo:2019gyi}). The square brackets denote the commutator. In addition, $\mathbb{M}_\textup{F}$ is the rotated neutrino mass matrix in the flavour basis,
\begin{equation}
    \mathbb{M}_\textup{F} = U 
    \begin{pmatrix}
        0 & 0 & 0 \\
        0 & \Delta m^2_{21} & 0 \\
        0 & 0 & \Delta m^2_{31}
    \end{pmatrix}
    U^\dagger,
\end{equation}
\noindent where \( \Delta m^2_{21} \) and \( \Delta m^2_{31} \) are the mass-squared differences between neutrino mass eigenstates, and \( U \) is the PMNS (Pontecorvo-Maki-Nakagawa-Sakata) mixing matrix. 

The key point for assessing the impact of NSI on $N_{\rm eff}$ is that the collisional integrals
\(\mathcal{I}(\varrho)\) depend on the couplings that appear in the
neutrino-electron scattering amplitudes. In the Standard Model, these
couplings are encoded in the matrices (in flavour space)
\begin{equation}
   G_L^{\text{SM}} = \mathrm{diag}\left(\tilde{g}_L,\; g_L,\; g_L\right),
   \qquad
   G_R^{\text{SM}} = \mathrm{diag}\left(g_R,\; g_R,\; g_R\right),
\end{equation}
where \(\tilde{g}_L = \sin^2{\theta_W}+1/2\) takes into account the charged-current contribution in the case of electron neutrinos. 
When NSI with electrons are present, these matrices are modified and they read 
\begin{equation}
   G_L \;=\;
   \begin{pmatrix}
      \tilde{g}_L+\varepsilon_{ee}^L & \varepsilon_{e\mu}^L & \varepsilon_{e\tau}^L \\[2pt]
      \varepsilon_{\mu e}^L & g_L+\varepsilon_{\mu\mu}^L & \varepsilon_{\mu\tau}^L \\[2pt]
      \varepsilon_{\tau e}^L & \varepsilon_{\tau\mu}^L & g_L+\varepsilon_{\tau\tau}^L
   \end{pmatrix},
   \ \ \ 
   G_R \;=\;
   \begin{pmatrix}
      g_R+\varepsilon_{ee}^R & \varepsilon_{e\mu}^R & \varepsilon_{e\tau}^R \\[2pt]
      \varepsilon_{\mu e}^R & g_R+\varepsilon_{\mu\mu}^R & \varepsilon_{\mu\tau}^R \\[2pt]
      \varepsilon_{\tau e}^R & \varepsilon_{\tau\mu}^R & g_R+\varepsilon_{\tau\tau}^R
   \end{pmatrix}.
   \label{eq:GLGR_NSI}
\end{equation}
\noindent
Since the collisional integrals \( \mathcal{I}(\varrho) \) involve products of the coupling matrices — one can find the full expressions in \cite{Bennett:2020zkv} — the presence of NSI leads to shifts in the corresponding Standard Model coefficients. In particular, the effective couplings entering the scattering terms are modified as follows:
\begin{align}
   g_L^2 &\;\longrightarrow\;
           \left(g_L+\varepsilon_{\alpha\alpha}^L\right)^{2}
           +\sum_{\beta\neq\alpha} \left|\varepsilon_{\alpha\beta}^L\right|^{2},
           \label{eq:gL2shift} \\[4pt]
   g_R^2 &\;\longrightarrow\;
           \left(g_R+\varepsilon_{\alpha\alpha}^R\right)^{2}
           +\sum_{\beta\neq\alpha} \left|\varepsilon_{\alpha\beta}^R\right|^{2},
           \label{eq:gR2shift} \\[4pt]
   g_L g_R &\;\longrightarrow\;
            \left(g_L+\varepsilon_{\alpha\alpha}^L\right)
            \left(g_R+\varepsilon_{\alpha\alpha}^R\right)
            + \sum_{\beta\neq\alpha} \mathrm{Re}\!\left[ \varepsilon^L_{\alpha\beta} \left(\varepsilon^R_{\alpha\beta}\right)^*\right].
            \label{eq:gLgRshift}
\end{align}

\noindent
These shifts modify the rate of neutrino interactions with the $e^\pm$ plasma, hence affecting the dynamics of the decoupling process. Consequently, both the decoupling temperature and \(N_{\text{eff}}\) become sensitive to the NSI parameters \(\varepsilon_{\alpha\beta}^{L,R}\).

Moreover, NSI also affect the matter-induced potentials responsible for flavour oscillations, and these modifications must be properly included in the evolution equation of the neutrino density matrix, specifically in the terms \( \mathbb{E}_\ell \) and \( \mathbb{P}_\ell \) of equation~\eqref{eq: matrixevol}. These contributions can be encoded in the matrix
\begin{equation}
    \mathbb{E}^{\rm NSI}_e =
    \begin{pmatrix}
    1 + \varepsilon_{ee}^{V} & \varepsilon_{e\mu}^{V} & \varepsilon_{e\tau}^{V} \\
    \varepsilon_{e\mu}^{V} & \varepsilon_{\mu\mu}^{V} & \varepsilon_{\mu\tau}^{V} \\
    \varepsilon_{e\tau}^{V} & \varepsilon_{\mu\tau}^{V} & \varepsilon_{\tau\tau}^{V}
    \end{pmatrix} \thinspace,
\end{equation}
so that the energy density and pressure contributions from the charged leptons read
\begin{equation}
    \mathbb{E}_\ell = \rho_e \, \mathbb{E}^{\rm NSI}_e, \qquad \mathbb{P}_\ell = P_e \, \mathbb{E}^{\rm NSI}_e\thinspace,
\end{equation}
where $\rho_e$ and $P_e$ are the electron energy density and pressure. These expressions contain the standard contribution (the unit term in the first element of $\mathbb{E}^{\rm NSI}_e$) and NSI corrections that depend on both the left- and right-handed couplings. Here, we have introduced the dimensionless parameter $\varepsilon_{\alpha\beta}^{V} \equiv \varepsilon_{\alpha\beta}^{R} + \varepsilon_{\alpha\beta}^{L}$, that is the vector coupling.


\subsection{Results for \texorpdfstring{\Neff}{Neff}}
\label{subsec:results_Neff}

In order to include the full set of relevant effects in the determination of \( N_{\rm eff} \), we perform a complete numerical computation using the publicly available code  \texttt{FortEPiaNO}\footnote{\url{https://bitbucket.org/ahep_cosmo/fortepiano_public}} (Fortran-Evolved Primordial Neutrino Oscillations) \cite{Bennett:2020zkv, Gariazzo:2019gyi}. The numerical setup is chosen to ensure high precision, obtaining \( N_{\rm eff} \) with a numerical error below \( 5 \times 10^{-4} \) in the standard scenario, a level of precision sufficient to probe the small effects introduced by non-standard interactions. To reach this accuracy, the absolute and relative tolerances of the differential equation solver are set to \( 10^{-6} \). The neutrino momentum integration is performed using Gauss-Laguerre quadrature with \( N_y = 50 \) momentum bins and a maximum comoving momentum \( y_{\text{max}} = 20 \). The initial conditions are set at \( x_{\text{in}} = 0.01 \), corresponding to a temperature much higher than the decoupling scale. We compute $\Neff$ from the final neutrino spectra, obtained at $x_\mathrm{fin} = 35$, which corresponds to a temperature $\simeq 0.015 \, \mathrm{MeV}$. We note that in the standard case (no NSI), the results from \texttt{FortEPiaNO} are in excellent agreement with other codes using a discrete momentum method to solve the quantum kinetic equation~\cite{Akita:2020szl,Froustey:2020mcq}. Other recent independent approaches have also been benchmarked against our standard results, including the adaptation of Direct Simulation Monte Carlo techniques to the early universe~\cite{Ovchynnikov:2024rfu,Ovchynnikov:2024xyd,Ihnatenko:2025kew} and the so-called ``Neural Boltzmann Equations''~\cite{Spinner:2026ytq}.

The results presented in this work for \( N_{\rm eff} \) are analogous to those reported in Ref.~\cite{deSalas:2021aeh}.
Here, we extend that analysis by exploring a wider region of the NSI parameter space, in order to obtain a more complete picture of their effects on BBN.
Although a large part of this region is already disfavoured by terrestrial experiments, this allows us to assess the sensitivity of neutrino decoupling and Big Bang Nucleosynthesis predictions to non-standard interactions in a complementary way
and to identify the regions compatible with current cosmological observables. 

\paragraph{One-parameter study}

\begin{figure}[t]
    \centering
    \includegraphics[width=\textwidth]{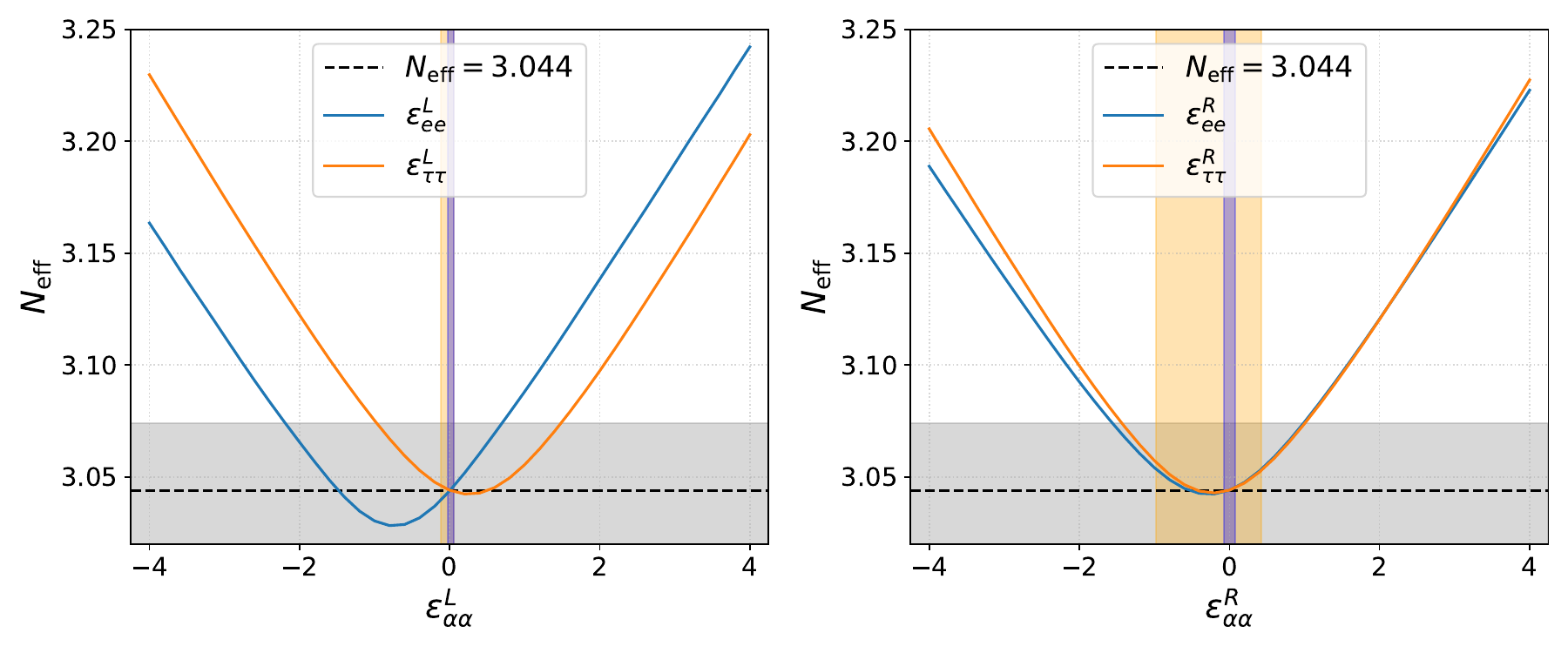}
    \caption{Values of \( N_{\rm eff} \) as a function of non-universal NSI parameters, \( \varepsilon^L_{\alpha\alpha} \) in the left panel and \( \varepsilon^R_{\alpha\alpha} \) in the right panel, for \(\alpha = \{e,\tau\}\). The dashed line corresponds to the standard prediction \( N_{\rm eff} = 3.044 \), and the shaded region corresponds to $\pm0.03$, which is the targeted 1$\sigma$ uncertainty from future cosmological observations \cite{CMB-S4:2026mge}. The shaded vertical bands correspond to the 90\% C.L. bounds shown in Table \ref{tab:NSIbounds}, in orange for the $\tau\tau$ coupling and in blue for the $ee$ ones.}
    \label{fig:fortepiano1}
\end{figure}

In Fig.~\ref{fig:fortepiano1}, we show the values of \( N_{\rm eff} \) as a function of \( \varepsilon_{\alpha\alpha}^{L} \) (left panel) and \( \varepsilon_{\alpha\alpha}^{R} \) (right panel), for \(\alpha = \{e,\tau\}\), varying one parameter at a time while keeping all others fixed to zero. For conciseness, we do not present the results for flavour-changing NSI parameters $\varepsilon_{\alpha \beta}$, which are gathered in Appendix~\ref{app:flavour_changing}. We do not include $\alpha = \mu$ in the results presented here, for two reasons. On the one hand, there are particularly stringent bounds from terrestrial experiments on NSI parameters involving muon neutrinos (see Table~\ref{tab:NSIbounds}). On the other hand, our analysis would yield almost exactly the same results as for the tau flavour. Indeed, in the universe at temperatures of MeV and below, the abundance of charged leptons other than electrons/positrons is negligible, such that muon and tau neutrinos have identical Standard Model interactions. Any change in the tau sector would therefore be equivalent in the muon sector, up to small differences due to their slightly different flavour mixings with electron neutrinos.

For the case of \( \varepsilon_{ee}^{L} \), the minimum of \( N_{\rm eff} \) occurs at \( \varepsilon_{ee}^{L} = -\tilde{g}_L \simeq -0.727 \)~\cite{Erler:2013xha}, as expected from the shift in the coupling structure.
In this configuration, the effective left-handed coupling to electrons vanishes, suppressing neutrino-electron interactions and leading to an earlier decoupling.
Similarly, for \( \varepsilon_{\tau\tau}^{L} \), the minimum is found at \( \varepsilon_{\tau\tau}^{L} = -g_L \simeq 0.273 \)~\cite{Erler:2013xha}, where the coupling to tau neutrinos cancels. An analogous argument applies to the right-handed couplings.
The minimum of \( N_{\rm eff} \) as a function of \( \varepsilon_{ee}^{R} \) occurs when this parameter cancels the Standard Model right-handed coupling, i.e., at \( \varepsilon_{ee}^{R} = -g_R \simeq - 0.233 \)~\cite{Erler:2013xha}. The same reasoning holds for \( \varepsilon_{\tau\tau}^{R} \), whose minimum appears at the same value.


%

\begin{figure}[!ht]
    \centering
    \includegraphics[width =0.95\textwidth]{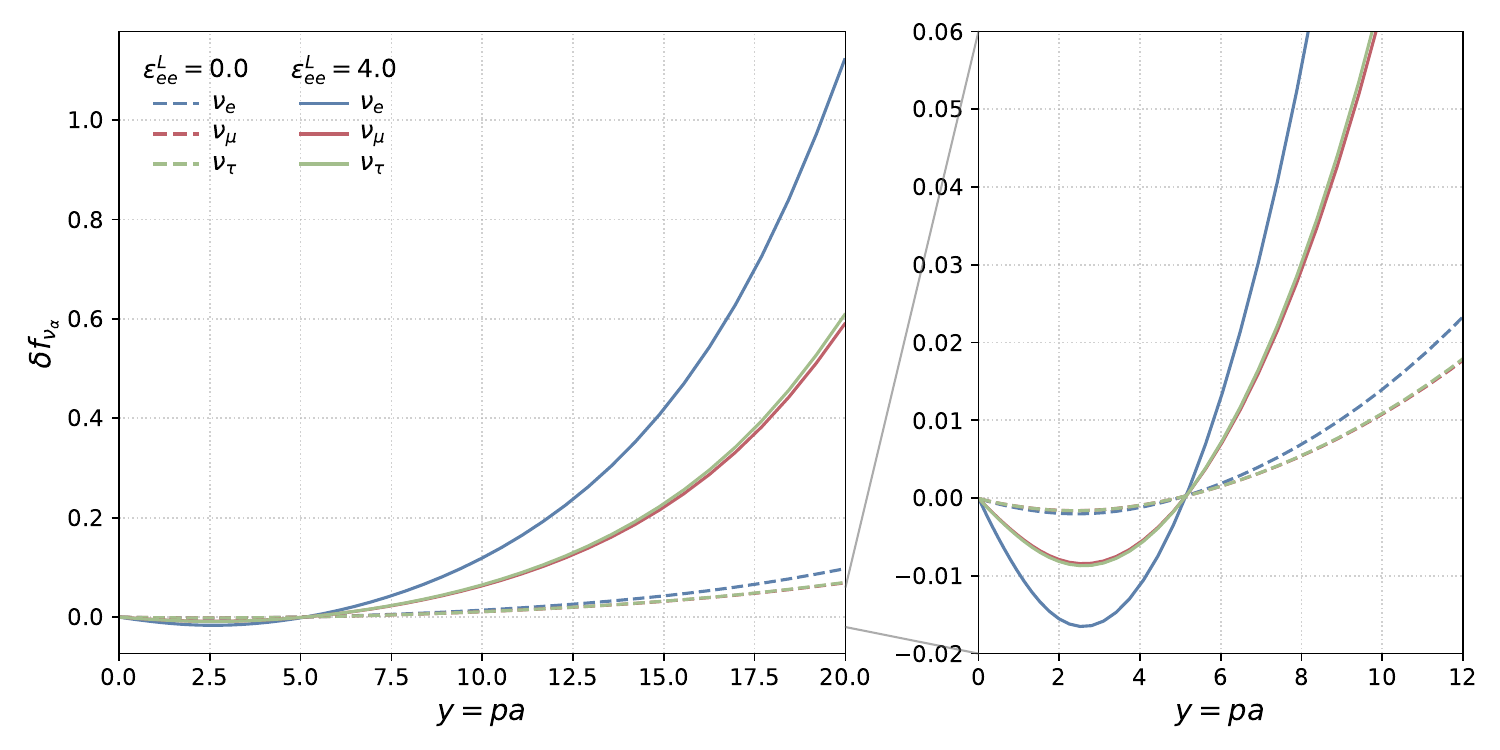}
    \caption{Non-thermal distortions of the frozen-out neutrino spectra [as defined in Eq.~\eqref{eq:def_distortions}], for the standard case ($\varepsilon_{ee}^L = 0$, dashed lines) and an extreme NSI case ($\varepsilon_{ee}^L = 4$, solid lines). The right panel is a zoomed-in version of the left panel.}
    \label{fig:fortepiano_distortions}
\end{figure}

As the non-standard couplings increase (in absolute value), neutrino decoupling is delayed and more entropy is transferred from $e^\pm$ annihilations into the neutrino sea, resulting in a larger $\Neff$. In addition, the neutrino distributions exhibit larger spectral distortions. In order to separate genuine distortions from an overall reheating of (anti)neutrinos, we define non-thermal distortions from the spectra after decoupling (i.e., at $x_\mathrm{fin}$) as
\begin{equation}
\label{eq:def_distortions}
    \varrho_{\alpha \alpha}(x_\mathrm{fin},y) \equiv \frac{1 + \delta f_{\nu_\alpha}(y)}{e^{y/z_\alpha} + 1} \, ,
\end{equation}
where the effective temperature $z_\alpha$ is chosen to reproduce the energy density of the actual distribution: $\frac{1}{2 \pi^2} \int_{0}^{\infty}{\mathrm{d} y \, y^3 \, \varrho_{\alpha \alpha}(x_\mathrm{fin},y)} = \frac78 \frac{\pi^2}{30} z_\alpha^4$. With this choice, $\delta f_{\nu_\alpha} = 0$ would correspond to a thermal Fermi-Dirac spectrum. These distortions are shown in Fig.~\ref{fig:fortepiano_distortions} for two runs with a different value of $\varepsilon_{ee}^L$: the standard scenario ($\varepsilon_{ee}^L = 0$), and the largest value we explored ($\varepsilon_{ee}^L = 4$). The zoomed-in version on the right panel of Fig.~\ref{fig:fortepiano_distortions} can be compared with Fig.~4 in~\cite{Froustey:2020mcq}, showing the consistency of the calculations in the standard case.

Consistent with the increase of $\Neff$ seen in Fig.~\ref{fig:fortepiano1}, the effective temperatures of neutrinos are largely increased between the two cases: for instance, $z_e(\varepsilon_{ee}^L = 0) \simeq 1.0017$ whereas $z_e(\varepsilon_{ee}^L = 4) \simeq 1.014$. In addition, there are significantly larger distortions with respect to the reference spectrum for $\varepsilon_{ee}^L = 4$.
These larger distortions will affect the weak rates determining the neutron/proton ratio during BBN, a visible effect in the synthesis of $^4\mathrm{He}$.

\paragraph{Two-parameter scan}

In order to investigate possible interference and combined effects among different NSI parameters, we allow for the simultaneous variation of two couplings. This multidimensional exploration is useful to identify degeneracies in the NSI parameter space that cannot be captured in one-parameter analyses.

\begin{figure}[t]
    \centering
    \includegraphics[width = 0.88\textwidth]{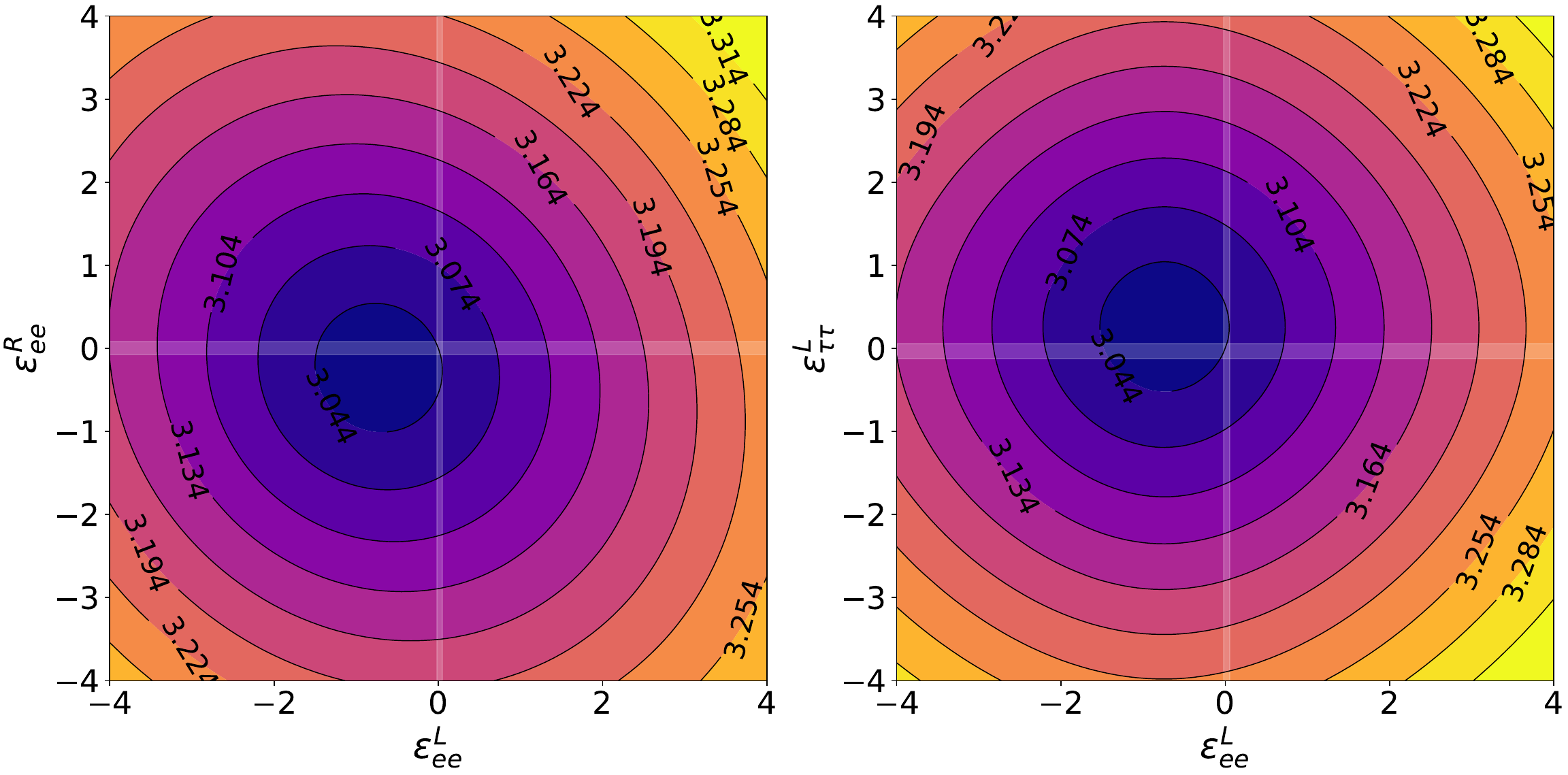}
    \caption{Values of \( N_{\rm eff} \) as a function of two diagonal NSI parameters. Left panel: \( \varepsilon^L_{ee} \) - \( \varepsilon^R_{ee} \); right panel: \( \varepsilon^L_{ee} \) - \( \varepsilon^L_{\tau\tau} \). White-shaded regions correspond to the 90\% C.L. experimental bounds obtained by varying one parameter at a time (Table \ref{tab:NSIbounds}).}
    \label{fig:fortepiano2}
\end{figure}

As an illustrative example, Fig.~\ref{fig:fortepiano2} shows contours of constant \(N_{\rm eff}\) in the plane spanned by different combinations of diagonal left- and right-handed NSI couplings. The resulting iso-\(N_{\rm eff}\) contours take an elliptical shape, as expected from the quadratic dependence of the effective neutrino--electron couplings on the NSI parameters, which also determines the location of the minima.

The ellipses of constant $\Neff$ on the left panel of Fig.~\ref{fig:fortepiano2}, centred on $(-\tilde{g}_L, -g_R)$, have principal axes rotated with respect to the coordinate axes. This ``tilt'' signals a cross term $\propto (\tilde{g}_L + \varepsilon_{ee}^L)(g_R + \varepsilon_{ee}^R)$ contributing to $\Neff$, which can be traced back to Eq.~\eqref{eq:gLgRshift}. This specific product enters the collision term, and thus the neutrino heating rate, because $m_e$ is non-negligible in the range of temperatures of interest~\cite{Escudero:2025kej}, all the more so for NSI that increase the effective couplings and delay neutrino decoupling to lower temperatures. Although not shown here, the same tilt of the iso-$\Neff$ ellipses can be seen in the $\{\varepsilon_{\tau \tau}^{L}, \varepsilon_{\tau \tau}^{R}\}$ plane. By contrast, there is no such cross term contribution in the plane $\{\varepsilon_{\alpha \alpha}^{L,R}, \varepsilon_{\beta \beta}^{L,R}\}$ with $\alpha \neq \beta$, as in the right panel of Fig.~\ref{fig:fortepiano2}.

Additional examples of the impact of non-standard neutrino--electron interactions on \(N_{\rm eff}\), for different choices of NSI parameters, are presented in \cite{deSalas:2021aeh}. 
In the present work, we limit ourselves to a representative subset, as our main focus is on the implications for Big Bang Nucleosynthesis observables.

\section{Consequences for Big Bang Nucleosynthesis}
\label{sec:results_BBN}

\subsection{Overview of BBN}

Following the decoupling of neutrinos and the rapid cooling of the early universe, conditions became suitable for the formation of light nuclei. This period, known as Big Bang Nucleosynthesis or primordial nucleosynthesis, marks a key moment in cosmic history: the genesis of the first elements. Occurring within the first few minutes after the Big Bang, BBN relies entirely on well-established SM physics and provides one of the most robust probes of the early universe. It predicts the primordial abundances of light elements such as deuterium (D or $^2$H), helium-3 ($^3$He), helium-4 ($^4$He), in remarkable agreement with the values inferred from astrophysical observations, see e.g.~\cite{Fields:2019pfx}; and lithium-7 ($^7$Li), which however does not match the observed value \cite{ParticleDataGroup:2026aaa}.
Hence, BBN plays a crucial role in constraining possible deviations from the standard cosmological scenario and probing new physics beyond the Standard Model.

Roughly one second after the Big Bang, when temperatures had dropped to the order of a few MeV, the particle content of the universe was limited to photons, (anti)neutrinos, electrons, positrons, and a small population of protons and neutrons, all in thermal equilibrium. Despite their low number density, nucleons play a central role during BBN, as they are the fundamental building blocks of the light nuclei that formed during this period. In particular, the neutron-to-proton ratio is a crucial quantity that determines the final elemental abundances, especially the production of helium-4. This ratio is governed by the following charged-current weak interaction processes:
\begin{align}\label{eq:weak1}
    \nu_e + n \ &\longleftrightarrow \ p + e^-, \\
    e^+ + n \ &\longleftrightarrow \ p + \bar{\nu}_e\thinspace, \\
    n \ &\longleftrightarrow \ p + e^- + \bar{\nu}_e\thinspace.
    \label{eq:weak3}
\end{align}
Here, \( n \) and \( p \) refer to neutrons and protons, respectively. If the rates for these interactions are rapid compared to the Hubble expansion rate, the system remains in chemical equilibrium.

After neutrino decoupling, the weak interaction rates involving neutrons and protons fall below the expansion rate of the universe, and chemical equilibrium can no longer be maintained. At temperatures below \( T \sim 0.7 \) MeV, the neutron-to-proton density ratio departs from its equilibrium value and  freezes out, with a residual decrease due to free neutron decays. This effectively sets the neutron abundance prior to the onset of nucleosynthesis.

After the first few minutes following the Big Bang, the baryonic composition of the universe was effectively established. Approximately 75\% of the baryonic mass remained in the form of hydrogen nuclei (protons), while nearly 25\% became bound in helium-4 nuclei. In addition to these dominant components, trace amounts of other light elements were also present, including deuterium, helium-3, and lithium-7. The observationally inferred primordial abundances used as reference in our analysis, namely helium-4 and deuterium, are summarised in Table~\ref{tab:primordial_abundances}. We perform a statistical analysis, comparing different primordial abundance determinations, in Section~\ref{subsec:statistical_analysis}. These values represent the current best estimates based on high-precision measurements and serve as crucial benchmarks for theoretical predictions from BBN computations. For helium-4 we adopt the recent Large Binocular Telescope (LBT) determination, since its smaller uncertainty allows us to assess the constraining power of helium in the parameter ranges explored here. Following the 
common convention adopted in cosmology, the abundances are 
given as the mass fraction \( Y_p \equiv 4n_{^4\rm He}/n_B \)
for helium-4, and the number density normalised to hydrogen
for deuterium,  \( ^2\mathrm{H}/\mathrm{H} \), where $\mathrm{H}\equiv n_H$, and $^2\mathrm{H}\equiv n_{^2\mathrm{H}}$. Given the lack of precision measurements (although recent progress was achieved in~\cite{Cooke:2026juw}) for ${}^3\mathrm{He}$, and the well-known ``lithium problem'', whose status is at present unclear~\cite{ParticleDataGroup:2026aaa}, we do not include other primordial abundance measurements beyond helium-4 and deuterium in our analysis.

In the framework of standard BBN, the theoretical predictions for the primordial abundances of light elements depend on well-established physics and inputs — such as the neutron lifetime $\tau_n$ and the relevant nuclear reaction rates — and there is only one free cosmological parameter: the baryon-to-photon ratio \( \eta_B \equiv n_B/n_\gamma \). This parameter controls the density of baryons available to form nuclei during BBN, and thereby influences the freeze-out and efficiency of the various nuclear reactions. It is often convenient to express it as \( \eta_{10} \equiv 10^{10} \eta_B \).

\begin{table}[t]
\centering
\setlength{\tabcolsep}{15pt}
\caption{Reference values for the primordial abundances of light elements adopted in this work. The helium-4 abundance is taken from the recent Large Binocular Telescope determination \cite{Aver:2026dxv}, whereas the deuterium abundance is taken from \cite{ParticleDataGroup:2026aaa}. Uncertainties correspond to the 1$\sigma$ level.}
\begin{tabular}{lc}
\toprule
\( Y_p \) & \( 0.2458 \pm 0.0013 \) \\[0.2em]
\( ^2\mathrm{H}/\mathrm{H} \times 10^5 \) & \( 2.508 \pm 0.029 \) \\[0.2em]
\bottomrule
\end{tabular}
\label{tab:primordial_abundances}
\end{table}

\subsection{Impact of relic neutrinos on Big Bang Nucleosynthesis}
\label{subsec:impact_neutrinos}

As BBN occurred right after neutrino decoupling, any modification in the decoupling process of relic neutrinos can alter the primordial abundances of light elements.

On the one hand, neutrinos contribute to the total radiation energy density, which determines the expansion rate of the universe during the BBN epoch. The effect of all relativistic species beyond photons is parametrised by \( N_{\rm eff} \). Therefore, any deviation from the standard value of this parameter alters the expansion rate and, consequently, the thermal history of the early universe — a mechanism commonly referred to as the \textit{background effect}. This change shifts the timing of the freeze-out of the weak interactions that interconvert neutrons and protons, thereby impacting the neutron-to-proton ratio and the final synthesis of light nuclei. In particular, a larger value of \( N_{\rm eff} \) leads to a faster expansion, which causes an earlier $n_n/n_p$ freeze-out. In addition, the faster expansion leaves less time for neutron decay before the deuterium bottleneck is overcome (a so-called ``clock effect''~\cite{Dodelson:1992km,Fields:1992zb,Froustey:2019owm}). All this results in a higher relic abundance of neutrons and ultimately enhances the production of \( ^4\mathrm{He} \). The deuterium abundance is in turn affected by the faster expansion, which shortens the duration of deuterium burning before freeze-out, resulting in a higher $^2\mathrm{H}$/H.

On the other hand, electron (anti)neutrinos participate directly in the charged-current weak interactions that govern the neutron-to-proton chemical equilibrium. Therefore, any deviation from the standard Fermi–Dirac spectra of neutrinos can modify the rates of these weak processes, altering the evolution of the neutron-to-proton ratio. Since this ratio sets the initial conditions for the synthesis of light elements, such distortions in the neutrino momentum distributions can lead to measurable changes in the primordial helium-4 mass fraction, as well as in the abundances of other light nuclei. One particular change of distributions can be an overall shift of temperature: if neutrinos are hotter, they shift the $n_n/n_p$ freeze-out to smaller temperatures, hence reducing the amount of neutrons at the onset of nucleosynthesis.

In order to translate these physical effects into bounds on NSI parameters, a fully numerical treatment of both neutrino decoupling and nuclear reaction networks is required. In this work, we adopt a two-step approach. 
First, we use the code \texttt{FortEPiaNO} to compute the evolution of the neutrino momentum distributions throughout the decoupling epoch for different values of the NSI parameters (some results being shown in Sec.~\ref{sec:results_Neff}). These time-dependent distributions consistently capture the impact of non-standard interactions on both neutrino spectra and weak interaction rates. 
The resulting neutrino distributions are then introduced into a modified version of the numerical code \texttt{PArthENoPE}\footnote{\url{https://parthenope.na.infn.it/}}~\cite{Pisanti:2007hk,Consiglio:2017pot,Gariazzo:2021iiu}, which allows us to compute the corresponding primordial light-element abundances. We have checked that we obtain the same qualitative variation of the abundances with NSI using the independent BBN code \texttt{PRIMAT}~\cite{Pitrou:2018cgg}, the exact values changing according to different choices of deuterium burning rates (see discussions in~\cite{Pitrou:2020etk,Pisanti:2020efz,Yeh:2020mgl,Pitrou:2021vqr}).

\subsection{Results for primordial abundances}

By varying only one NSI parameter at a time, we generate predictions for \( Y_p \) and \( ^2\rm H/H \). The resulting curves for the non-universal NSI parameters are shown in Fig.~\ref{fig:BBN_1D}, while those for flavour-changing couplings are gathered in Appendix~\ref{app:flavour_changing}. Throughout this analysis, we assume $\eta_{10} = 6.115$ (best fit from \emph{Planck} \cite{Aghanim:2018eyx, ParticleDataGroup:2026aaa}) and a neutron lifetime $\tau_n = 879.4 \, \mathrm{s}$~\cite{Gariazzo:2021iiu}. The observationally allowed regions for \(Y_p\) and \(^2{\rm H}/{\rm H}\) are shown in Fig.~\ref{fig:BBN_1D} as shaded bands, using the \(1\sigma\) intervals listed in Table~\ref{tab:primordial_abundances}. It is however delicate to directly extract $1\sigma$ constraints on the NSI parameters from these plots, and we defer a complete statistical analysis to section~\ref{subsec:statistical_analysis}.

 \begin{figure}[!ht]
    \centering
    \includegraphics[width=0.95\textwidth]{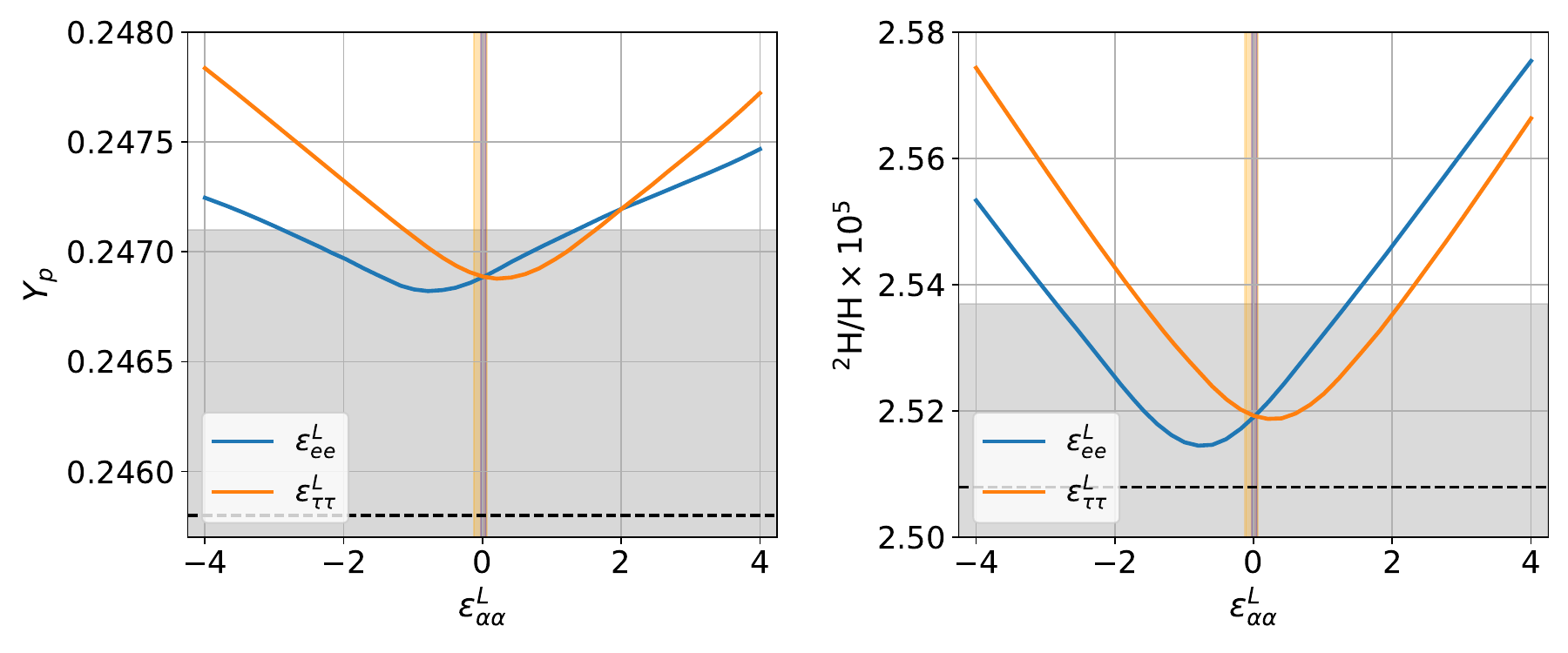}
    \includegraphics[width=0.95\textwidth]{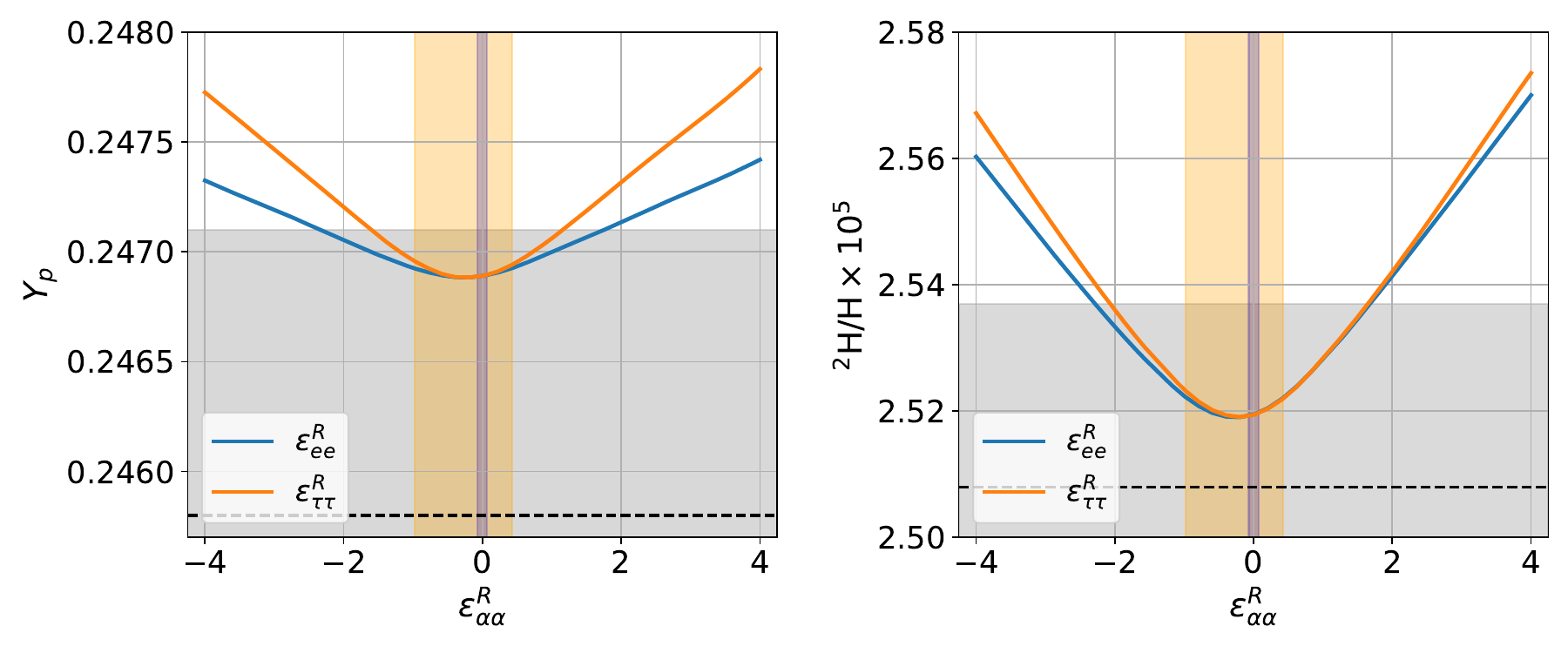}
    \caption{Primordial abundances of helium-4 (mass fraction) and deuterium (number density normalised to hydrogen) as functions of the NSI parameters: non-universal left-handed  \( \varepsilon_{ee}^{L} \) and \( \varepsilon_{\tau\tau}^{L} \) (top panels), non-universal right-handed \( \varepsilon_{ee}^{R} \) and \( \varepsilon_{\tau\tau}^{R}\) (lower panels).
    The horizontal dashed lines represent the observational central values, while the shaded grey regions correspond to the $1\sigma$ intervals (both listed in Table \ref{tab:primordial_abundances}).
    The shaded vertical bands correspond to the 90\% C.L. bounds shown in Table \ref{tab:NSIbounds}.}
    \label{fig:BBN_1D}
\end{figure}

\begin{figure}[!ht]
    \centering
    \includegraphics[width=\textwidth]{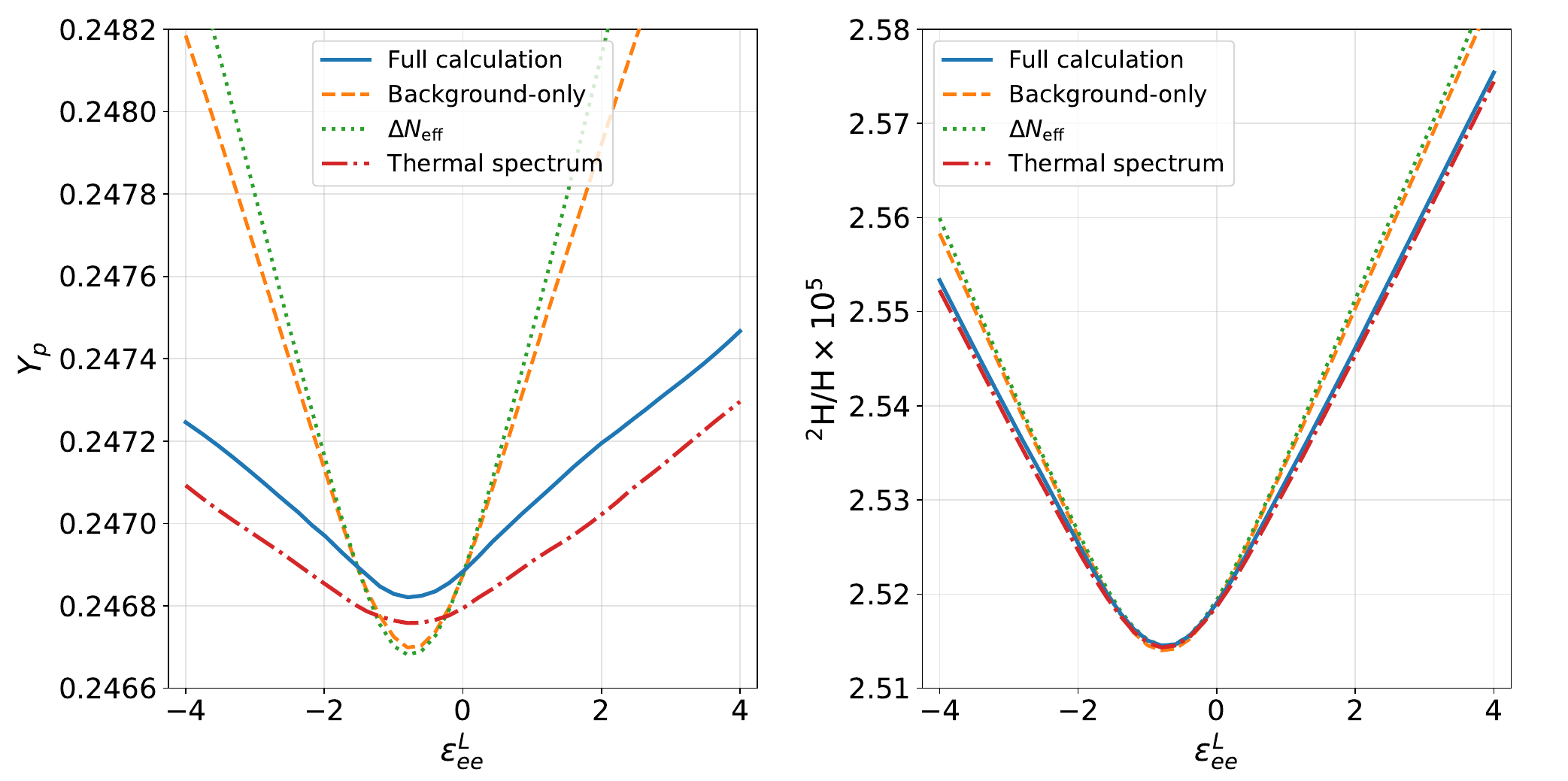}
    \caption{Primordial helium mass fraction $Y_p$ (\emph{left}) and deuterium abundance $\mathrm{^2H/H}$ (\emph{right}) as functions of $\varepsilon^L_{ee}$. We compare different calculations: the full one (blue solid line, identical to Fig.~\ref{fig:BBN_1D}), the ``background-only'' case (orange dashed line), a constant $\Delta N_{\rm eff}$ shift (green dotted line), and a calculation using a thermal $\nu_e$ spectrum with the same energy density as the actual distribution at each time (red dash-dotted line). See text for details.}
    \label{fig:comparison}
\end{figure}

The primordial abundances show a broad dependence on the NSI parameters similar to that found for $N_{\rm eff}$ in Fig.~\ref{fig:fortepiano1}, reflecting the important role of the modified neutrino energy density during BBN. However, this correspondence is not exact, since modifications of the $\nu_e$ and $\bar\nu_e$ distributions also affect the neutron-to-proton conversion rates, to which $Y_p$ is particularly sensitive. To disentangle these effects, we show in Fig.~\ref{fig:comparison} a detailed comparison for the representative case $\varepsilon^L_{ee}$.

Besides the full computation, Fig.~\ref{fig:comparison} includes three different scenarios\footnote{We emphasise that the extra scenarios are inconsistent in the physics they describe, but they allow us to clearly identify the mechanisms at play.}. First, in the ``background-only'' case, we retain the same modifications to the cosmological background as in the full calculation, while switching off the NSI-induced modifications of the weak rates, which are kept at their $\varepsilon=0$ values. This preserves the full time-dependent background evolution obtained with \fortepiano, including the changes in the neutrino energy density. In the ``$\Delta N_{\rm eff}$'' case, instead, we retain the standard neutrino distributions and weak rates and add an extra contribution to the energy density, corresponding to the value of $\Neff$ after decoupling. In other words, we add the $\Delta \Neff$ values shown in Fig.~\ref{fig:fortepiano1} (which could equivalently be taken from \cite{deSalas:2021aeh}) to the standard \texttt{PArthENoPE} engine. Finally, we perform a calculation (``thermal spectrum'') where the actual distributions are replaced by thermal Fermi-Dirac ones, specifically Eq.~\eqref{eq:def_distortions} with $\delta f_{\nu_e} = 0$, but calculated for each $x$. This removes the non-thermal spectral distortions while preserving the $\nu_e$ energy-density evolution, allowing us to isolate the impact of the spectral shape on the weak rates (in a similar fashion to the approach in~\cite{Froustey:2019owm,Froustey:2020mcq}).

The background-only and $\Delta N_{\rm eff}$ results are very close over the range shown, and are nearly indistinguishable for deuterium. The changes in abundances can be understood from the effect of the modified radiation density discussed in Sec.~\ref{subsec:impact_neutrinos}. An increase in $N_{\rm eff}$ enhances the expansion rate, leading to a larger helium abundance through its effect on the neutron-to-proton evolution ($n/p$ freeze-out at higher temperature and clock effect on neutron decay), and to a
larger deuterium abundance by reducing the time available for deuterium burning. Thus, both abundances generally increase in regions where the NSI produce a larger radiation density. Furthermore, the closeness of the background-only and $\Delta \Neff$ results shows that the time variation of $\Delta\Neff$, from 0 to its asymptotic value used as input in the $\Delta \Neff$ calculation, has a subdominant effect. This is consistent with the larger discrepancy between the orange and green curves in the $Y_p$ panel than the $^2\mathrm{H/H}$ one, since the clock effect affecting deuterium burning takes place once $\Neff$ has reached its asymptotic value.

The remaining difference between the full calculation and the background-only prediction is therefore associated with the modification of the weak rates through the $\nu_e$ and $\bar\nu_e$ distributions\footnote{There is no visible difference between the background-only and full calculations at $\varepsilon = 0$ because, when we do not read an external neutrino distribution table in \texttt{PArthENoPE}, it uses by default hard-coded results corresponding to the standard $\varepsilon = 0$ case.}. Crucially, simply changing $\Neff$ yields quantitatively wrong $Y_p$ predictions, because it ignores the fact that the change in $\Neff$ is due to changes in the neutrino spectra, and not to some extra-ultrarelativistic degree of freedom.

For helium, the effect of the modified weak rates is clearly visible. This is expected, since the helium mass fraction is particularly sensitive to the weak rates governing neutron--proton interconversion, which determine the $n/p$ ratio before the onset of efficient nucleosynthesis [Eqs.~\eqref{eq:weak1}--\eqref{eq:weak3}]. Therefore, modifications in the electron-neutrino and antineutrino spectra induced by NSI can produce deviations from the pure \(N_{\rm eff}\)-driven behaviour. The calculation using an energy-matched thermal spectrum follows a very similar dependence on $\varepsilon^L_{ee}$ as the full result. This shows that it captures the ``moderation'' of $Y_p$ variations compared to the background-only effect: for instance, at large $\varepsilon_{ee}^L$, neutrinos have higher energies, which increases the $n \leftrightarrow p$ rates and delays the $n/p$ freeze-out, resulting in a smaller helium abundance. This effect, in the opposite direction to the variations of $\Neff$, must be included to obtain accurate predictions. We note that the thermal spectrum results are systematically shifted towards lower values of $Y_p$. Since the two calculations share the same background evolution and the same $\nu_e$ energy density at each
time, their difference isolates the effect of the non-thermal spectral shape. Qualitatively, enforcing a thermal shape redistributes part of the energy contained in the high-momentum tail (see Fig.~\ref{fig:fortepiano_distortions}) towards intermediate momenta, in the region relevant for the neutron--proton conversion integrals. This increases the weak rates, delaying $n/p$ freeze-out and thus slightly lowering $Y_p$.

By contrast, the primordial abundance of deuterium is much less sensitive to these spectral effects. The full, background-only, $\Delta N_{\rm eff}$, and thermal spectrum predictions remain much closer to one another, showing that the $^2\mathrm{H/H}$ dependence on the NSI parameter is predominantly controlled by the modified expansion history. Small differences become visible only towards the edges of the range shown, indicating a subdominant sensitivity to the NSI-induced changes in the neutrino distributions.

\begin{figure}[!ht]
    \centering
    \includegraphics[width=0.44\textwidth]{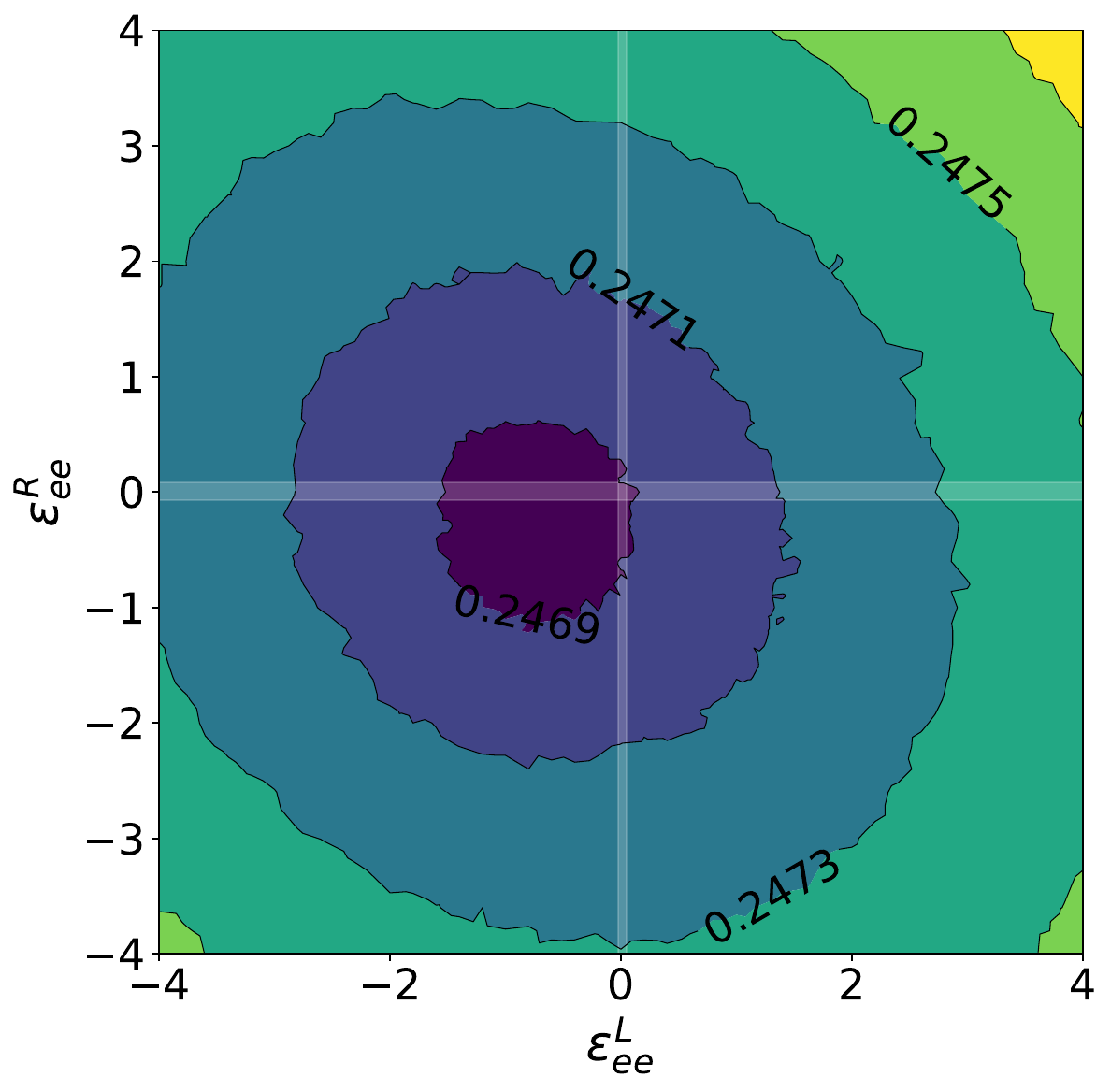}
    \includegraphics[width=0.44\textwidth]{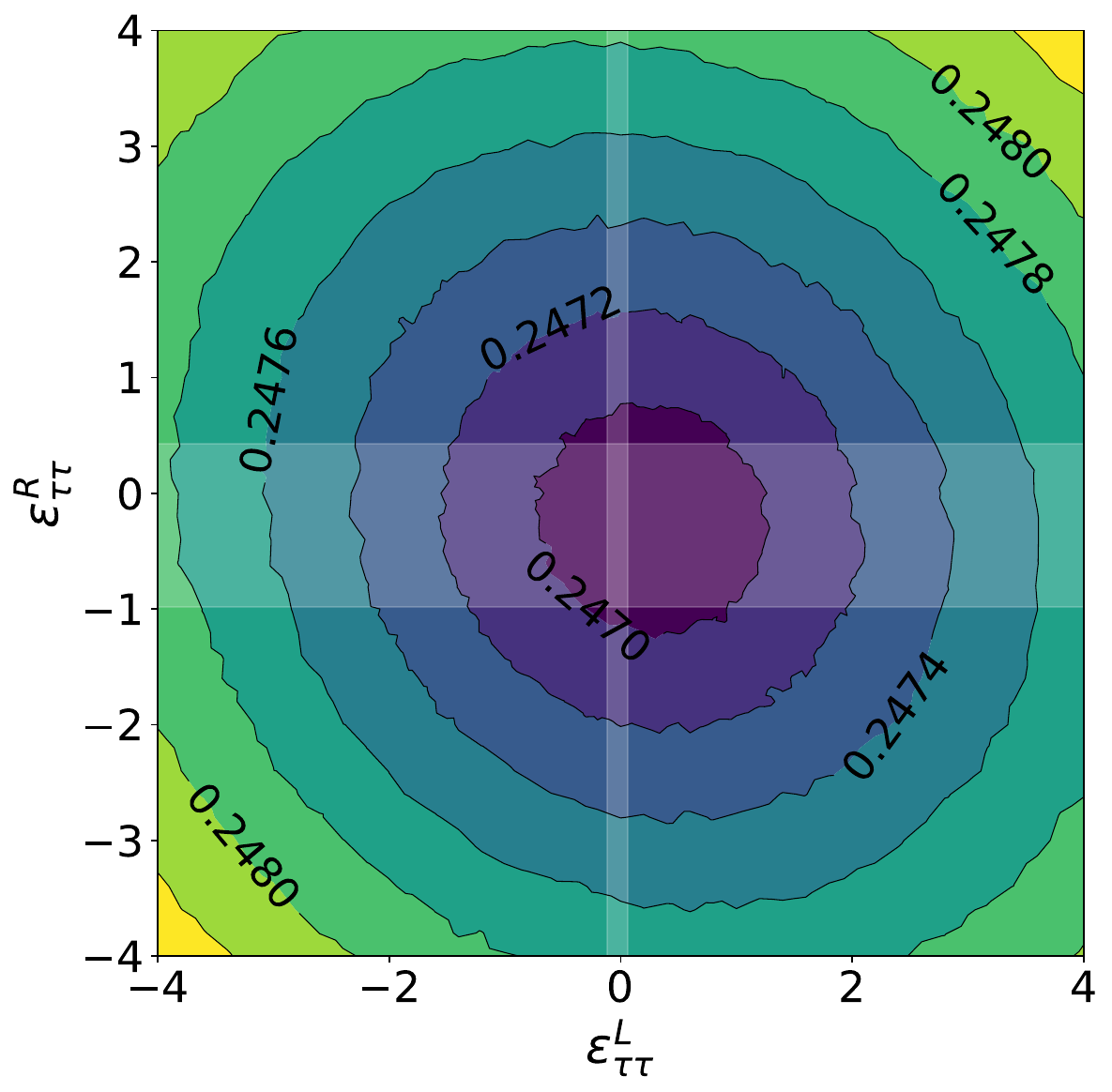}\\[0.5em]
    \includegraphics[width=0.44\textwidth]{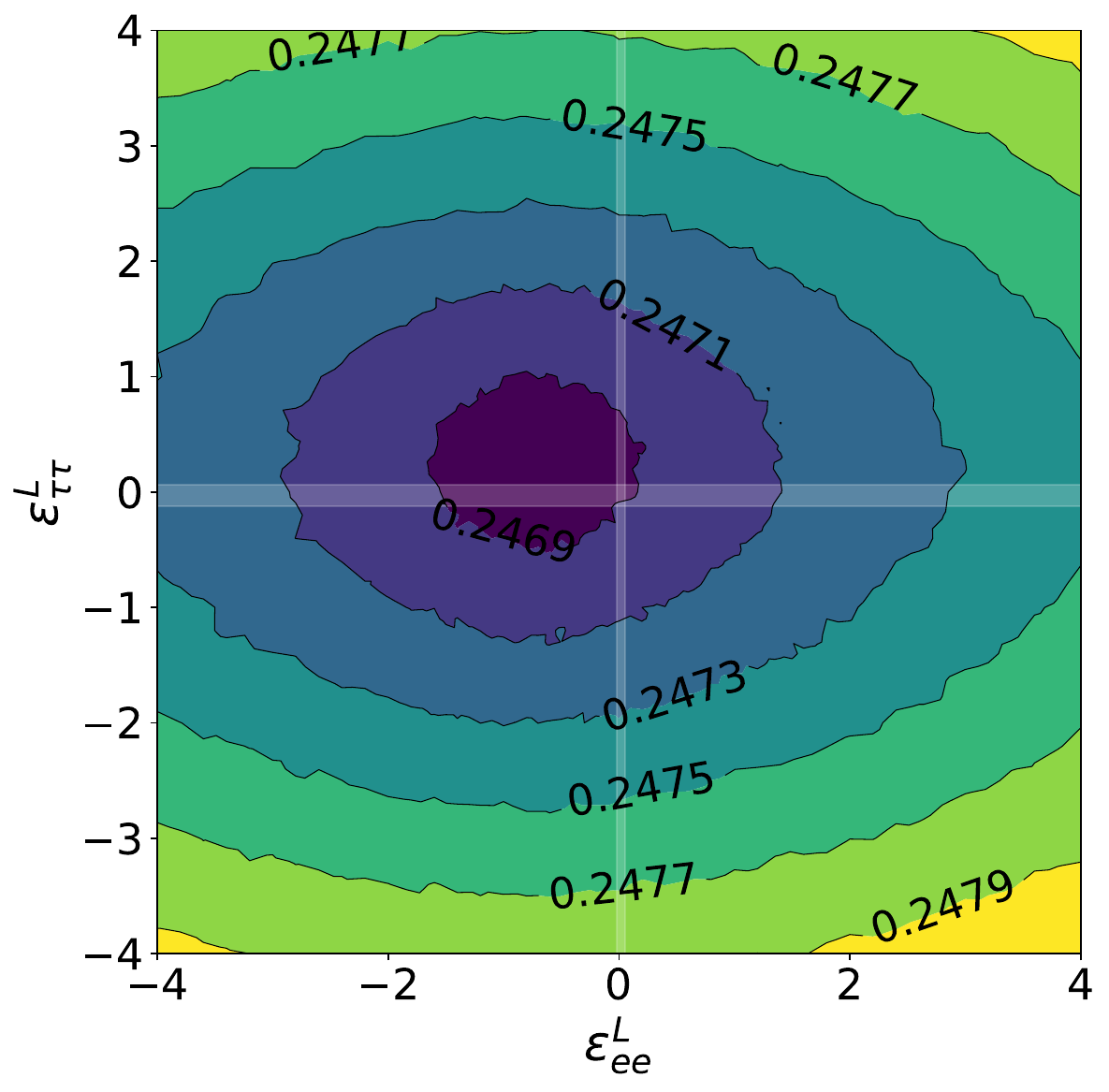}
    \includegraphics[width=0.44\textwidth]{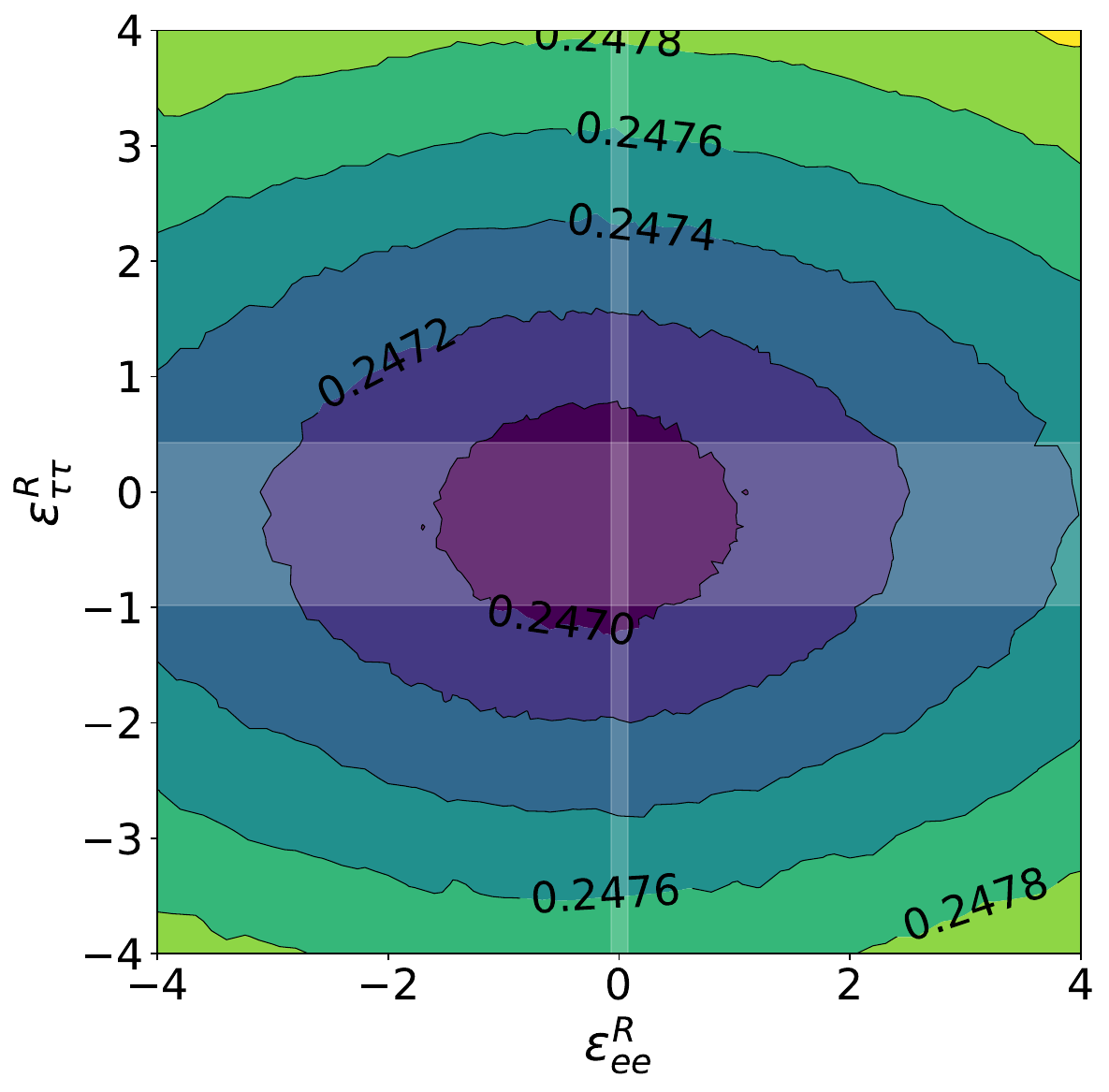}
    \caption{Values of \( Y_p \) when varying two diagonal NSI parameters simultaneously. Top left panel: \( \varepsilon^L_{ee} \) - \( \varepsilon^R_{ee} \); top right panel: \( \varepsilon^L_{\tau\tau} \) - \( \varepsilon^R_{\tau\tau} \); bottom left panel: \( \varepsilon^L_{ee} \) - \( \varepsilon^L_{\tau\tau} \); bottom right panel: \( \varepsilon^R_{ee} \) - \( \varepsilon^R_{\tau\tau} \). White-shaded regions correspond to the 90\% C.L. experimental bounds obtained by varying one parameter at a time (Table \ref{tab:NSIbounds}).}
    \label{fig:BBN_Yp_2D}
\end{figure}

\begin{figure}[!ht]
    \centering
    \includegraphics[width=0.44\textwidth]{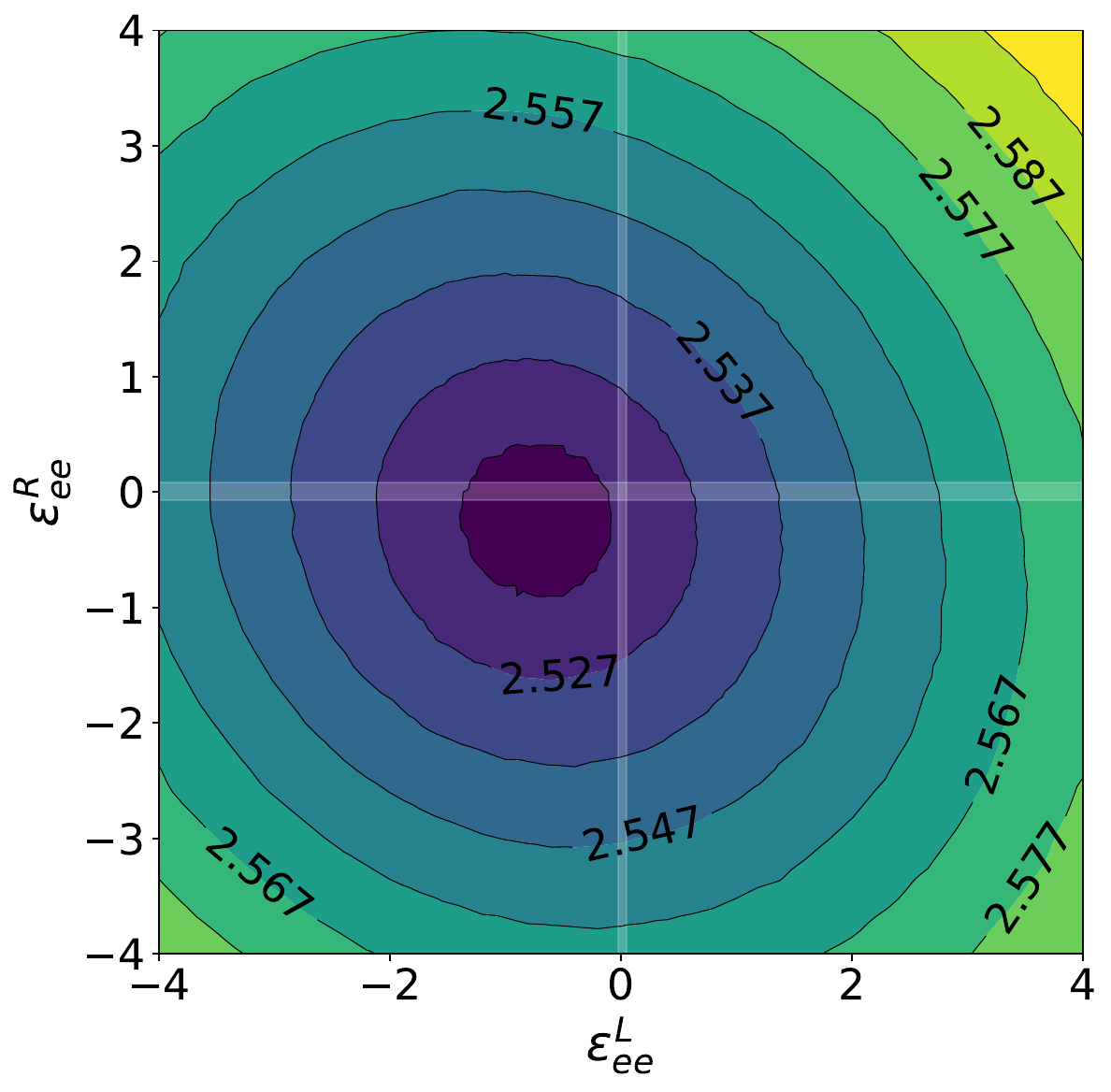}
    \includegraphics[width=0.44\textwidth]{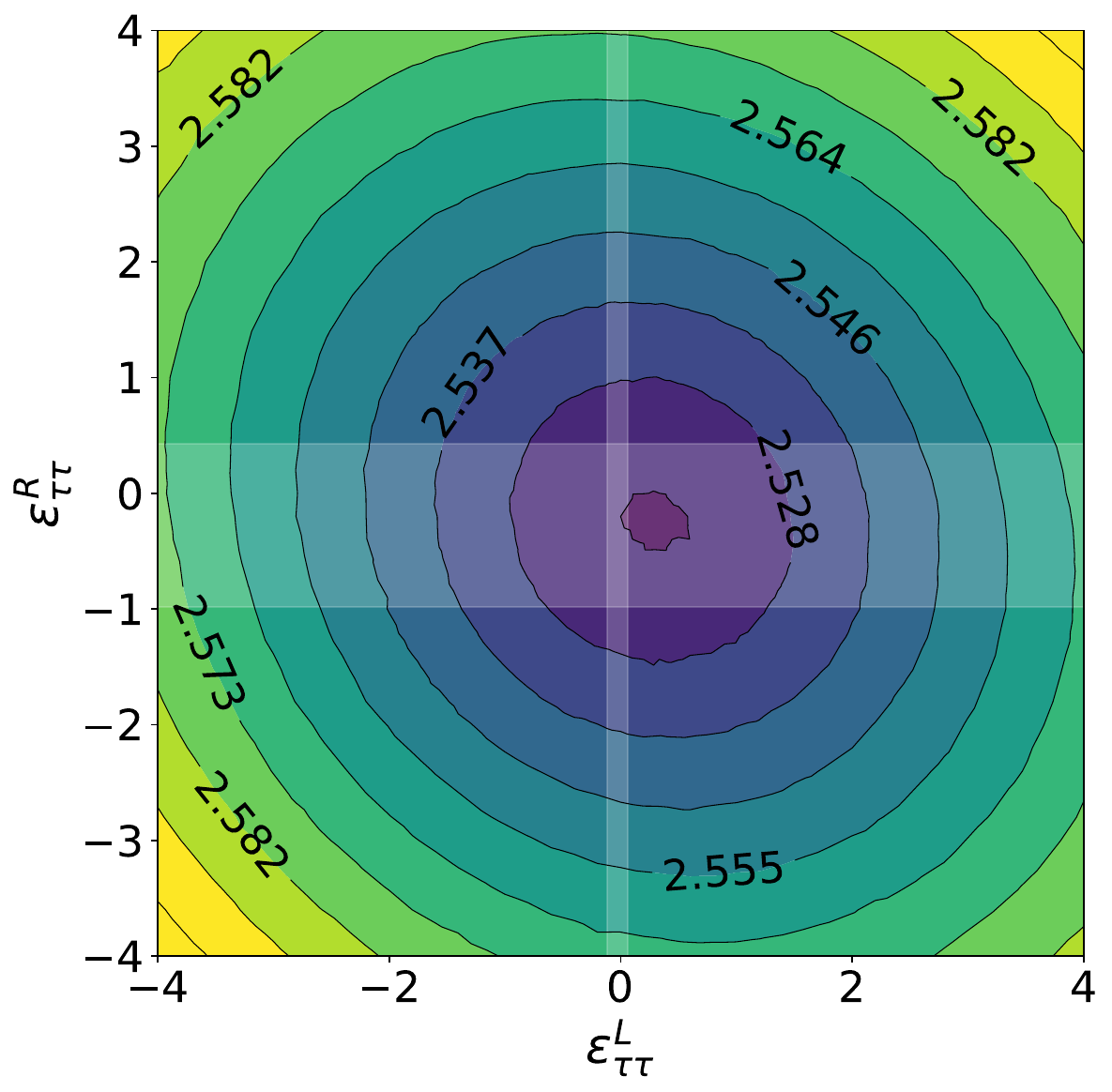}\\[0.5em]
    \includegraphics[width=0.44\textwidth]{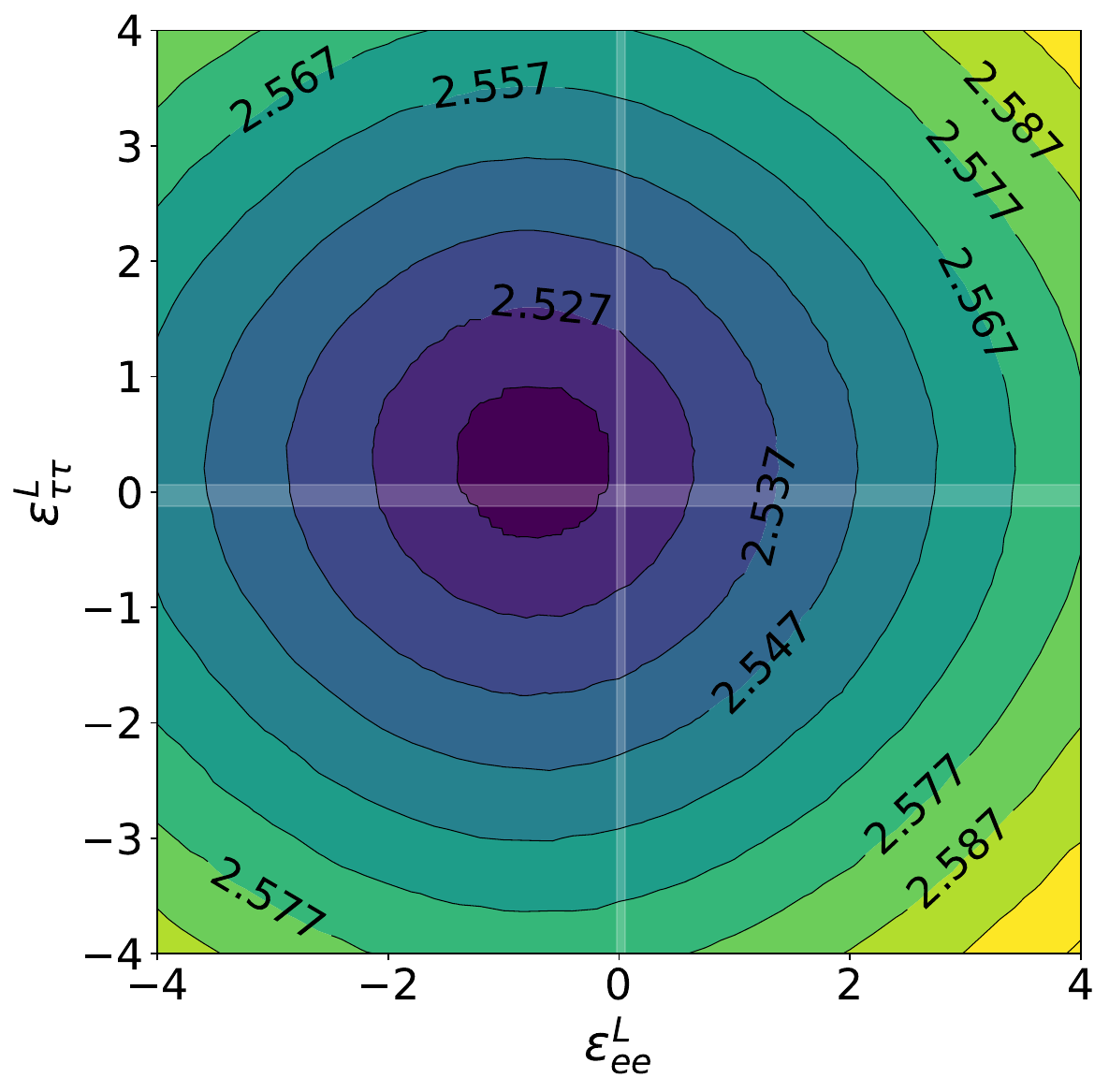}
    \includegraphics[width=0.44\textwidth]{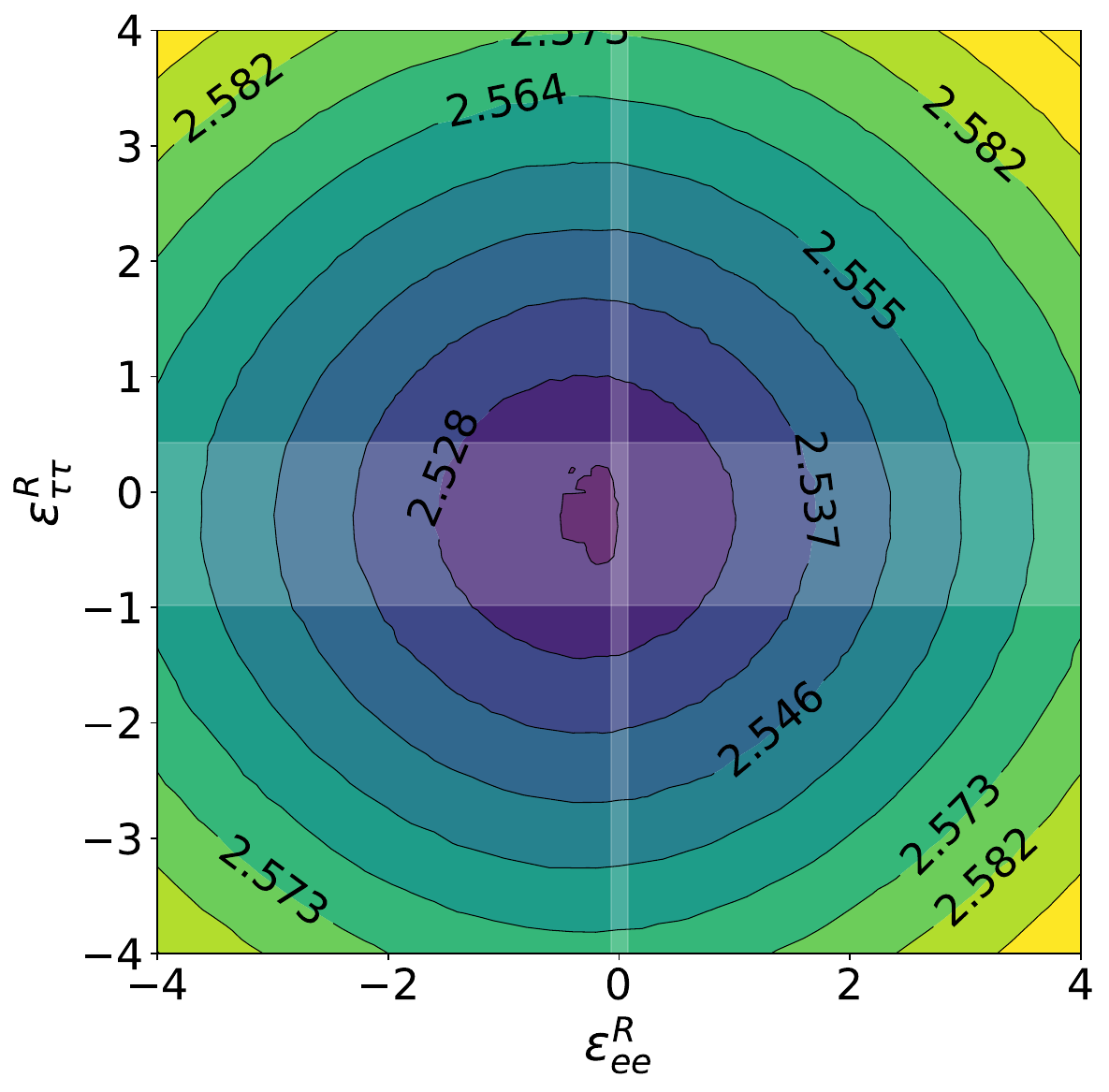}
    \caption{Same as Fig.~\ref{fig:BBN_Yp_2D}, but for the primordial deuterium abundance, \(^2\mathrm{H}/\mathrm{H}\times10^5\).}
    \label{fig:BBN_DH_2D}
\end{figure}

The conclusions drawn from the one-dimensional analysis are further supported by the two-dimensional scans. The $Y_p$ and $^2\mathrm{H}/\mathrm{H}$ contours shown in Figs.~\ref{fig:BBN_Yp_2D} and~\ref{fig:BBN_DH_2D}, respectively, display the same broad structure as the corresponding $N_{\rm eff}$ maps shown in Fig.~\ref{fig:fortepiano2}, reflecting their common dependence on the NSI-induced modifications of the neutrino energy density. The variation of
$Y_p$ across the parameter space is comparatively mild, whereas deuterium shows a more pronounced variation and follows the $N_{\rm eff}$ pattern more closely. Overall, the two-dimensional results confirm the physical picture emerging from the one-dimensional analysis.

\subsection{Statistical analysis}
\label{subsec:statistical_analysis}

We now quantify the constraints that can be set on NSI parameters from cosmology alone, comparing various measurements of the relevant observables.

\paragraph{Datasets and likelihood}

\renewcommand{\arraystretch}{1.2}

When varying one (or two) NSI parameter(s), denoted as $\varepsilon$, the pipeline presented in the previous sections allows us to obtain the predictions $\Neff(\varepsilon)$, $Y_p(\varepsilon)$ and $^2\mathrm{H/H}(\varepsilon)$. Given the experimental measurements of these quantities, we combine Gaussian likelihoods and calculate $\chi^2 = - 2 \ln \mathcal{L}$ as
\begin{equation}
\label{eq:chi2}
\begin{aligned}
    \chi^2(\varepsilon) &= \chi^2_{\Neff}(\varepsilon) + \chi^2_{Y_p}(\varepsilon) + \chi^2_{^2\mathrm{H/H}}(\varepsilon) \\
    &= \left(\frac{\Neff(\varepsilon) - \hat{N}_\mathrm{eff}}{\sigma_{\Neff}}\right)^2 + \left(\frac{Y_p(\varepsilon) - \hat{Y}_p}{\sigma_{Y_p}}\right)^2 + \left(\frac{^2\mathrm{H/H}(\varepsilon) - \widehat{^2\mathrm{H/H}}}{\sigma_{^2\mathrm{H/H}}}\right)^2 \, ,
\end{aligned}
\end{equation}
where the central values are denoted with a hat and gathered in Table~\ref{tab:dataset_combinations}, and uncertainties are the uncorrelated sum of the theoretical and observational errors, $\sigma_X^2 = (\sigma_X^\mathrm{th})^2 + (\sigma_X^\mathrm{obs})^2$, the latter being given in Table~\ref{tab:dataset_combinations}. The errors coming from the numerical predictions are $\sigma_{\Neff}^\mathrm{th} = 0.0005$~\cite{Bennett:2020zkv}, $\sigma_{Y_p}^\mathrm{th} = 0.00015$~\cite{Gariazzo:2021iiu} and $\sigma_{^2\mathrm{H/H} \times 10^5}^\mathrm{th}=0.06$~\cite{Pisanti:2020efz} (we consider that the baryon density is known exactly).

We use the experimental values obtained from different dataset combinations, in order to assess the relative contributions of $\Neff$ and the primordial abundances, and to forecast the future potential of cosmological measurements. We first consider the best data available at the time of the previous $\Neff$-focused study~\cite{deSalas:2021aeh}, namely the $Y_p$ measurement by Aver \emph{et al.}~\cite{Aver:2020fon}, the deuterium one by Cooke \emph{et al.}~\cite{Cooke:2017cwo}, and the CMB observations by \emph{Planck}~\cite{Aghanim:2018eyx}. This combination is dubbed ``2021 status''. Since then, improvements were made in the measurement of $Y_p$ with the most recent LBT determination~\cite{Aver:2026dxv} and several deuterium measurements~\cite{Kislitsyn:2024jvk,Guarneri:2024qxi}, leading to a new PDG-recommended value~\cite{ParticleDataGroup:2026aaa}. Using these updated measurements corresponds to the ``Current BBN + \emph{Planck}'' combination. We further study the impact of different $Y_p$ determinations in Appendix~\ref{app:helium4}. On the CMB side, a recent combination of SPT-3G, \emph{Planck} and ACT observations (``SPA'') in~\cite{SPT-3G:2025bzu} has provided a tighter constraint on $\Neff$, which we use in the ``Current BBN + SPA'' combination. We note two difficulties with this constraint: it prefers a value $\Neff < 3$, invariably in tension with our results (see Fig.~\ref{fig:fortepiano1}), while an independent analysis using similar datasets reports a larger central value $\hat{N}_\mathrm{eff} = 3.031$~\cite{Tristram:2025you}. Finally, we estimate the future constraining power of CMB observations with the expected sensitivity of CMB-S4~\cite{CMB-S4:2026mge}, in the combination ``Current BBN + CMB forecast''.

\begin{table}[!t]
\centering
\caption{Experimental determinations of the observables used for our statistical analysis. From left to right, the $\Neff$ constraints are obtained from \emph{Planck}~\cite{Aghanim:2018eyx}, the combination of SPT, ACT and \emph{Planck}~\cite{SPT-3G:2025bzu}, and a forecast of CMB-S4 capabilities~\cite{CMB-S4:2026mge}. The ``old'' helium-4 abundance determination is taken from~\cite{Aver:2020fon}, the more recent one being the LBT determination~\cite{Aver:2026dxv}; for deuterium the 2021 status uses the result from~\cite{Cooke:2017cwo}, while our ``current BBN'' values come from the PDG recommendation~\cite{ParticleDataGroup:2026aaa}. Uncertainties correspond to the 1$\sigma$ level.}
\setlength{\tabcolsep}{2.8pt}
\footnotesize
\begin{tabular}{rcccc}
\toprule
 & 2021 status & Current BBN + \emph{Planck} & Current BBN + SPA & Current BBN + CMB forecast  \\ \midrule 
$\Neff$ & $2.99 \pm 0.17$ & $2.99 \pm 0.17$ & $2.82 \pm 0.12$ & $3.044 \pm 0.030$ \\
$Y_p$ &  $0.2453 \pm 0.0034$ & $0.2458 \pm 0.0013$ & $0.2458 \pm 0.0013$ & $0.2458 \pm 0.0013$ \\
$^2\mathrm{H}/\mathrm{H} \! \times \! 10^5$ & $2.527 \pm 0.030$ & $2.508 \pm 0.029$ & $2.508 \pm 0.029$ & $2.508 \pm 0.029$ \\
\bottomrule
\end{tabular}
\label{tab:dataset_combinations}
\end{table}

\paragraph{One-parameter scan results}

\begin{figure}[!ht]
    \centering
    \includegraphics[width=\textwidth]{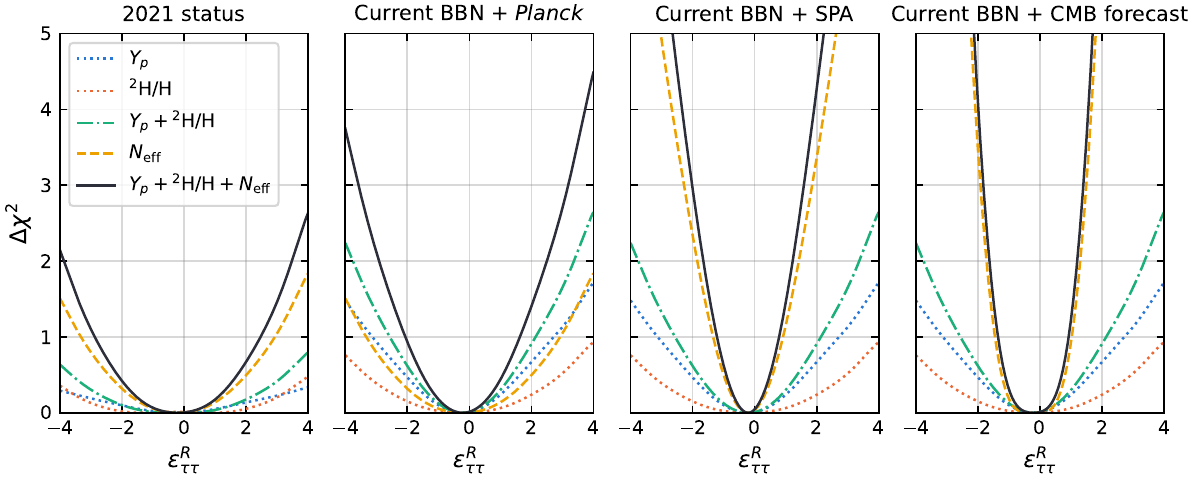}
    \caption{Profile likelihood of $\varepsilon_{\tau \tau}^R$ for the different dataset combinations presented in Table~\ref{tab:dataset_combinations}. We plot $\Delta \chi^2 \equiv \chi^2 - \chi^2_\mathrm{min}$, and show the contributions of BBN measurements (dotted and dash-dotted lines), $\Neff$ (dashed orange line) and the full combination (solid black line).}
    \label{fig:PL_GR33}
\end{figure}

We illustrate the role of the different observables by showing in Fig.~\ref{fig:PL_GR33} the profile likelihood from Eq.~\eqref{eq:chi2} in one example, namely for $\varepsilon = \varepsilon_{\tau \tau}^R$ (cf. right panel of Fig.~\ref{fig:fortepiano1} and second row of Fig.~\ref{fig:BBN_1D}).  
The constraints coming from the recent, sub-percent measurements of the primordial abundances are somewhat tighter than the ones stemming from the \emph{Planck} $\Neff$ measurement. However, when combining with the SPA $\Neff$ measurement, it is the latter that dominates the likelihood. But this happens at the expense of the goodness of fit: in this example, the $\chi^2$ at the best-fit point goes from 0.81 (Current BBN + \emph{Planck}) to 4.85 (Current BBN + SPA). This reflects the $\approx 2 \sigma$ tension between $\Neff = 2.82 \pm 0.12$ and the standard value, which cannot be absorbed by any value of $\varepsilon$ since $\Neff \geq 3.04$ throughout this scan. The tighter interval obtained with SPA is therefore driven by that offset rather than by improved agreement. With the CMB forecast, centred on the Standard Model value, essentially all the constraining power comes from $\Neff$ with no degradation of the goodness of fit ($\chi^2_\mathrm{min} = 0.71$).

We perform a Bayesian analysis by computing $\mathbb{P}(\varepsilon | D) \propto \exp[-\chi^2(\varepsilon)/2]$, $D$ denoting the data, with a flat prior on $\varepsilon \in [-4,4]$. The results are summarised in Table~\ref{tab:statistical_results}. We report 68~\% highest posterior density credible intervals, because 90 \% intervals are too prior-sensitive to be quoted with confidence. In order to compute these intervals, we interpolate between our discrete $\varepsilon$ data points with a degree-8 polynomial. Comparison with the laboratory constraints in Table~\ref{tab:NSIbounds} demonstrates that cosmology is not competitive, but still provides a completely independent probe of the same physics in a vastly different environment.

\begin{table}[!ht]
\centering
\caption{Results of our one-parameter statistical analysis for the ``Current BBN + \emph{Planck}'' and ``Current BBN + SPA'' combinations, using a flat prior on $\varepsilon \in [-4,4]$. We provide 68 \% C.I. and $\chi^2$ of the highest likelihood point. The high values of $\chi^2_\mathrm{min}$ when using the SPA $\Neff$ constraints stem from the large $\Neff$ values throughout our parameter space compared to the smaller than 3 preferred value—this tension is the main driver of the tighter C.I. with SPA.}
\setlength{\tabcolsep}{15pt}
\small
\begin{tabular}{ccccc}
\toprule
 & \multicolumn{2}{c}{Current BBN + \emph{Planck}} & \multicolumn{2}{c}{Current BBN + SPA}  \\
 & 68 \% C.I. & $\chi^2_\mathrm{min}$ & 68 \% C.I. & $\chi^2_\mathrm{min}$ \\ \midrule
$\varepsilon_{ee}^L$ & $[-2.71, 1.23]$ & $0.68$ & $[-1.83 , 0.35]$ & $4.28$ \\
$\varepsilon_{ee}^R$ & $[-2.33, 1.72]$ & $0.81$ & $[-1.42 , 0.86]$ & $4.83$ \\
$\varepsilon_{\tau \tau}^L$ & $[-1.56, 2.07]$ & $0.81$ & $[-0.80 , 1.31]$ & $4.83$ \\
$\varepsilon_{\tau \tau}^R$ & $[-2.03, 1.60]$ & $0.81$ & $[-1.27 , 0.84]$ & $4.85$ \\
$\varepsilon_{e \tau}^L$ & $[-1.42, 1.42]$ & $0.85$ & $[-0.78 , 0.78]$ & $4.98$ \\
$\varepsilon_{e \tau}^R$ & $[-1.42,1.42]$ & $0.85$ & $[-0.78, 0.78]$ & $4.98$ \\
\bottomrule
\end{tabular}
\label{tab:statistical_results}
\end{table}

\begin{figure}[!ht]
    \centering
    \includegraphics[width=\textwidth]{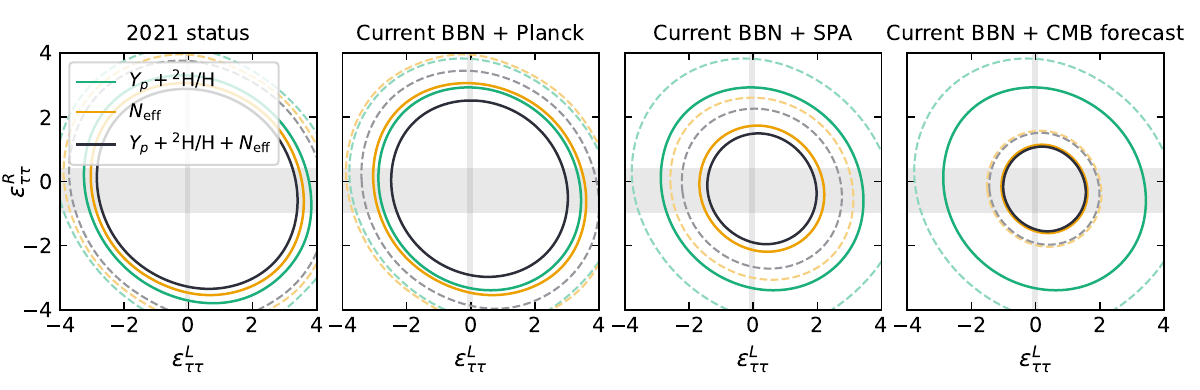}
    \caption{Posterior distributions in the $\{\varepsilon_{\tau \tau}^L, \varepsilon_{\tau \tau}^R\}$ plane. The solid and dashed contours contain 68~\% and 90~\% of the posterior probability, respectively. Grey bands show the 90 \% C.L. constraints from laboratory experiments (see Table~\ref{tab:NSIbounds}). Each panel corresponds to a different choice of dataset combination, and each colour corresponds to different likelihood combinations: BBN only (green), CMB $\Neff$ only (orange), and BBN + CMB (black).}
    \label{fig:Bayesian_2D_GL33_GR33}
\end{figure}

\paragraph{Two-parameter scans} Within our framework, we can study the correlations between different NSI parameters, as shown in Figs.~\ref{fig:fortepiano2}, \ref{fig:BBN_Yp_2D} and \ref{fig:BBN_DH_2D}. Since the primordial abundances follow the same trend as the $\Neff$ results, with the correlations discussed in Sec.~\ref{subsec:results_Neff}, we expect preferred regions following the same patterns. We confirm this by performing the same statistical analysis as above, the $\varepsilon$ variable being now a set of two NSI coefficients. We show one example of such a study, where we simultaneously vary $\varepsilon_{\tau \tau}^L$ and $ \varepsilon_{\tau \tau}^R$, in Fig.~\ref{fig:Bayesian_2D_GL33_GR33}.

It is clear that current constraints are not competitive compared with the one-parameter 90 \% C.L. bounds from terrestrial experiments, shown in grey bands. However, this comparison can be misleading: we do not know what correlations may exist between the $\varepsilon$ NSI coefficients in laboratory experiments, such that the allowed ranges may be significantly larger, and along with it the comparative constraining power of cosmology.

\section{Conclusions}
\label{sec:conclusion}

In this work, we have explored some cosmological implications of neutrino non-standard interactions with electrons, focusing on their impact during two key stages in the early universe: neutrino decoupling and Big Bang Nucleosynthesis. These epochs are particularly sensitive to the properties of neutrinos and thus offer a unique opportunity to probe possible deviations from Standard Model predictions.

As a first step, we studied the effect of non-standard interactions on neutrino decoupling, reproducing the results obtained in Ref.~\cite{deSalas:2021aeh} for the effective number of relativistic species \(N_{\rm eff}\). 
We extended this analysis by exploring a wider region of the NSI parameter space, confirming the robustness of the numerical framework and using \(N_{\rm eff}\) as a diagnostic quantity to characterise the impact of NSI on the thermal history of the early universe.

We then computed the impact of neutrino NSI with electrons on Big Bang Nucleosynthesis, focusing on the primordial helium mass fraction \(Y_p\) and the deuterium-to-hydrogen ratio \(^2\mathrm{H}/\mathrm{H}\). The BBN response cannot be reduced to the corresponding change in $N_{\rm eff}$, since this change is associated with modifications of the (anti)neutrino distributions. More precisely, the modification of the cosmological background contributes to the overall behaviour of both abundances and largely determines the deuterium response. However, the helium abundance receives an important additional contribution from the modified electron (anti)neutrino distributions through the neutron--proton conversion rates, leading to visible departures from a purely \Neff-driven behaviour.

We have also performed a statistical analysis combining different measurements of $N_{\rm eff}$ and the primordial abundances (see Table~\ref{tab:statistical_results}). The constraining power of these observables depends on the adopted datasets, but the resulting cosmological bounds remain weaker than the corresponding one-parameter constraints from terrestrial experiments. Nevertheless, they provide an independent probe of neutrino interactions at vastly different energy and time scales.

Finally, we note that the strength of the BBN bounds is directly tied to the primordial abundance determinations used in the analysis. In particular, the recent LBT measurement of $Y_p$, owing to its small uncertainty, provides significantly stronger helium-based constraints than less precise determinations. Our comparison with alternative helium measurements also illustrates that apparently stronger constraints may arise from tensions between the measured central value and the theoretical predictions rather than from improved precision. 
Future improvements in the precision and robustness of primordial abundance measurements, for both deuterium and helium, as well as in CMB observations, could further enhance the sensitivity of cosmological probes to non-standard neutrino interactions. 
In this context, cosmology will continue to play an important role in testing the fundamental properties of neutrinos and possible extensions of the Standard Model.

\acknowledgments

At IFIC Valencia, this work was supported by the Spanish grants PID2023-147306NB-I00 and CEX2023-001292-S (MCIU/AEI/10.13039/501100011033). J.M.\ is supported by the grant FPU24/00735. S.G.\ also acknowledges financial support through the Ram\'on y Cajal contract RYC2023-044611-I funded by MICIU/AEI/10.13039/501100011033 and FSE+. S.G. and O.P. are supported by the Research grant TAsP (Theoretical Astroparticle Physics) funded by Istituto Nazionale di Fisica Nucleare (INFN). O.P. is also supported by Ministero dell’Universit\`a e della Ricerca (MUR), PRIN2022 program (Grant PANTHEON 2022E2J4RK), Italy.

\appendix

\section{Cosmological observables for flavour-changing NSI}
\label{app:flavour_changing}

\begin{figure}[!ht]
    \centering
    \includegraphics[width=0.45\textwidth]{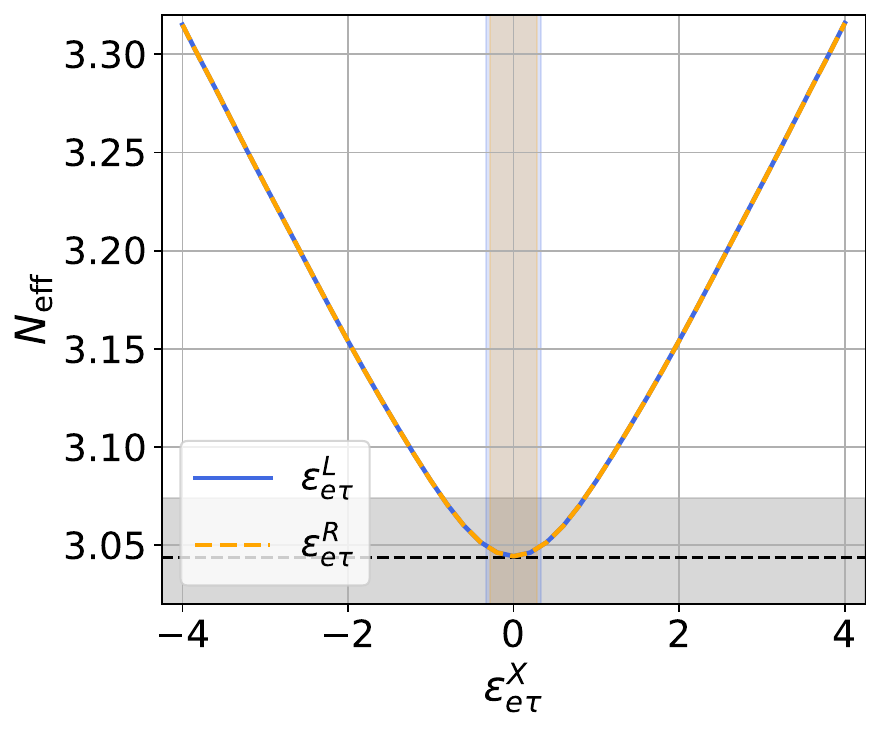}
    \includegraphics[width=0.95\textwidth]{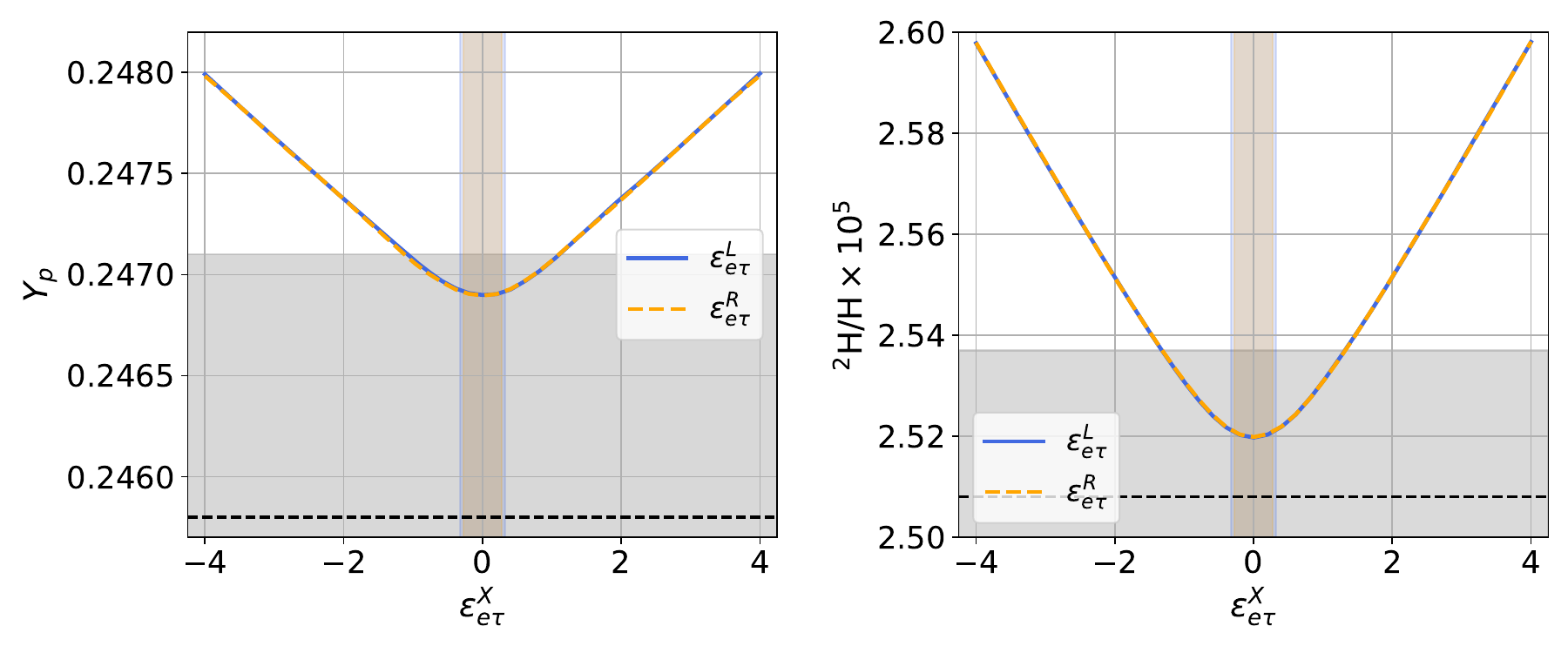}
    \caption{$N_{\rm eff}$ (top), $Y_p$ (bottom left), and $\mathrm{^2 H/H}$ (bottom right) as functions of the flavour-changing NSI parameters $\varepsilon^X_{e\tau}$. Blue solid and orange dashed lines correspond to $\varepsilon^L_{e\tau}$ and $\varepsilon^R_{e\tau}$, respectively. In the top panel, the horizontal dashed line marks $N_{\rm eff}=3.044$, while the grey band shows the targeted $1\sigma$ uncertainty, $\sigma(N_{\rm eff})=0.03$, from future cosmological observations \cite{CMB-S4:2026mge}. In the bottom left and bottom right panels, the horizontal dashed lines and grey bands indicate the observational central values and $1\sigma$ intervals listed in Table~\ref{tab:primordial_abundances}. The vertical shaded bands indicate the 90\% C.L. bounds reported in Table~\ref{tab:NSIbounds}.}
    \label{fig:BBN_1D_offdiag}
\end{figure}

For completeness, in this Appendix we present the corresponding cosmological predictions for flavour-changing NSI, which were omitted from the main text for clarity. Figure~\ref{fig:BBN_1D_offdiag} shows $N_{\rm eff}$, $Y_p$, and $^2\mathrm{H}/\mathrm{H}$ as functions of the representative flavour-changing couplings $\varepsilon^L_{e\tau}$ and $\varepsilon^R_{e\tau}$.

When one flavour-changing NSI parameter is varied at a time, the dependence of the cosmological observables is almost exactly symmetric around \(\varepsilon_{e\tau}^{L,R}=0\). This follows from the fact that the off-diagonal couplings enter the collision terms quadratically [see Eqs.~\eqref{eq:gL2shift}--\eqref{eq:gLgRshift}]. Consequently, $N_{\rm eff}$ and the primordial abundances increase symmetrically as the magnitude of the flavour-changing parameter grows. Moreover, the predictions obtained by varying $\varepsilon^L_{e\tau}$ or $\varepsilon^R_{e\tau}$ are practically indistinguishable for all three observables, with their differences remaining below the estimated numerical uncertainty of the corresponding codes.

The behaviour of the primordial abundances broadly follows that of $N_{\rm eff}$. As in the diagonal case discussed in the main text, deuterium tracks the modification of the radiation energy density particularly closely, whereas $Y_p$ also retains an additional sensitivity to the modified $\nu_e$ and $\bar\nu_e$ distributions entering the neutron--proton conversion rates.

\begin{figure}
    \centering
    \includegraphics[width=0.44\textwidth]{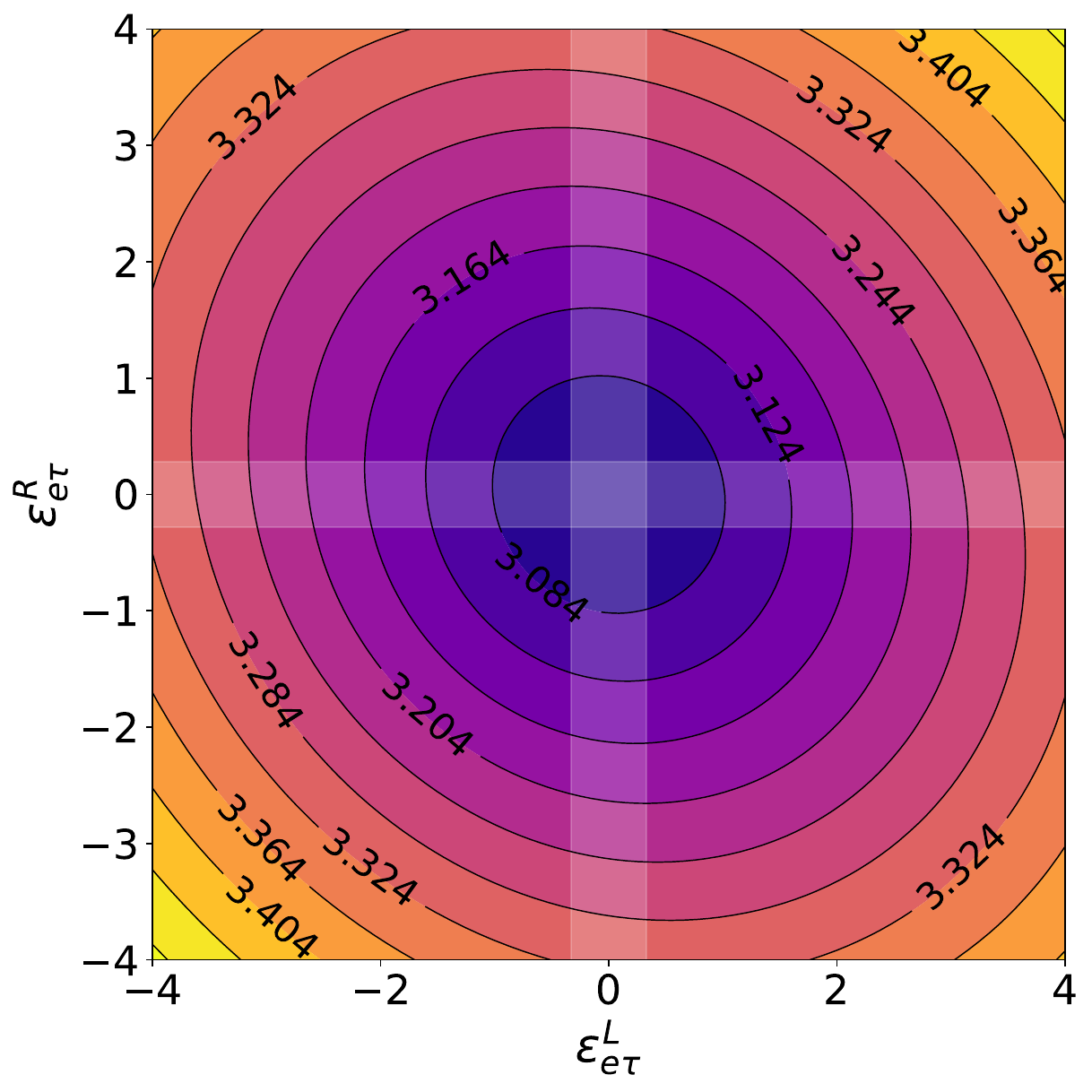}\\[0.5em]
    \includegraphics[width=0.44\textwidth]{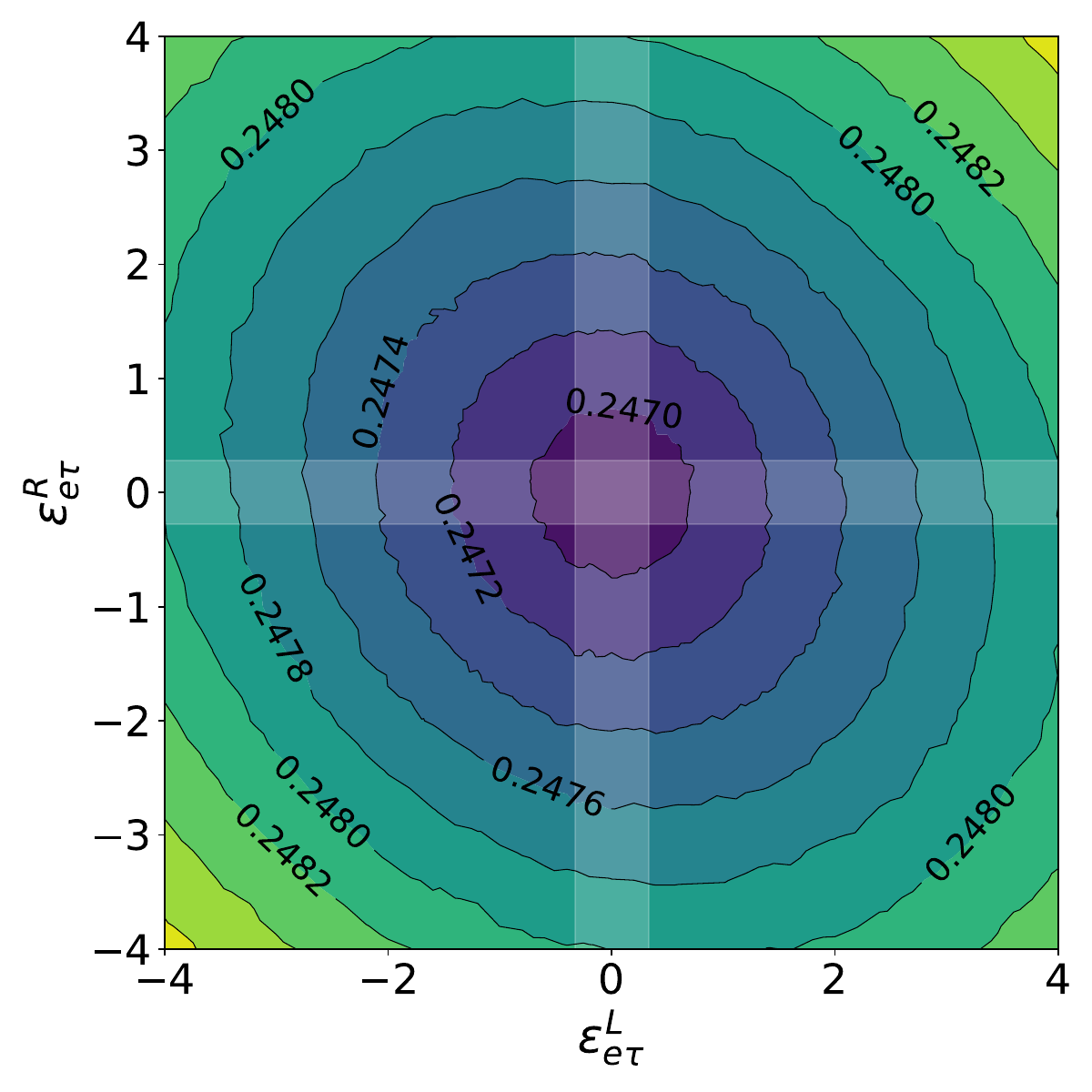}
    \includegraphics[width=0.44\textwidth]{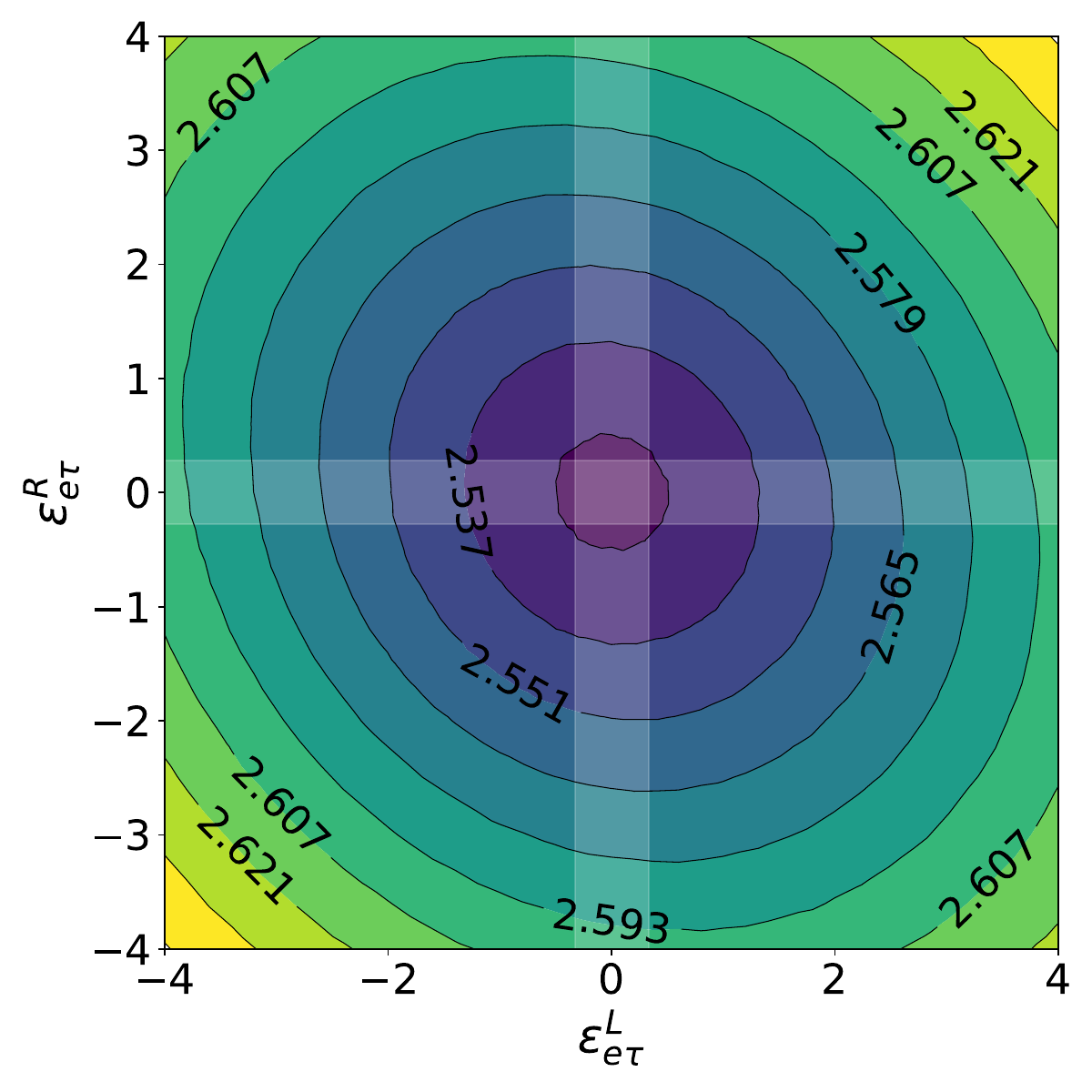}
    \caption{Values of $\Neff$ (top), \( Y_p \) (bottom left) and \(^2\mathrm{H}/\mathrm{H}\times10^5\) (bottom right) obtained by simultaneously varying the flavour-changing NSI parameters \( \varepsilon^L_{e\tau} \) and \( \varepsilon^R_{e\tau} \). White-shaded regions correspond to the 90\% C.L. experimental bounds obtained by varying one parameter at a time (Table \ref{tab:NSIbounds}).}
    \label{fig:BBN_offdiag_2D}
\end{figure}

Figure~\ref{fig:BBN_offdiag_2D} shows the simultaneous variation of $\varepsilon^L_{e\tau}$ and $\varepsilon^R_{e\tau}$. The same overall correspondence between $N_{\rm eff}$ and the primordial abundances is visible in the two-dimensional plane. In this case, however, the dependence is not only determined by the individual quadratic terms, since the collision integrals also contain mixed left--right contributions [see Eq.~\eqref{eq:gLgRshift}]. This is the reason for the ``tilt'' of the isocontours in Fig.~\ref{fig:BBN_offdiag_2D}. This highlights the information on the interplay between two flavour-changing couplings that is absent from one-dimensional scans.

\newpage

\section{Impact of helium-4 measurement}
\label{app:helium4}

\begin{figure}[!ht]
    \centering
    \includegraphics[width=0.95\textwidth]{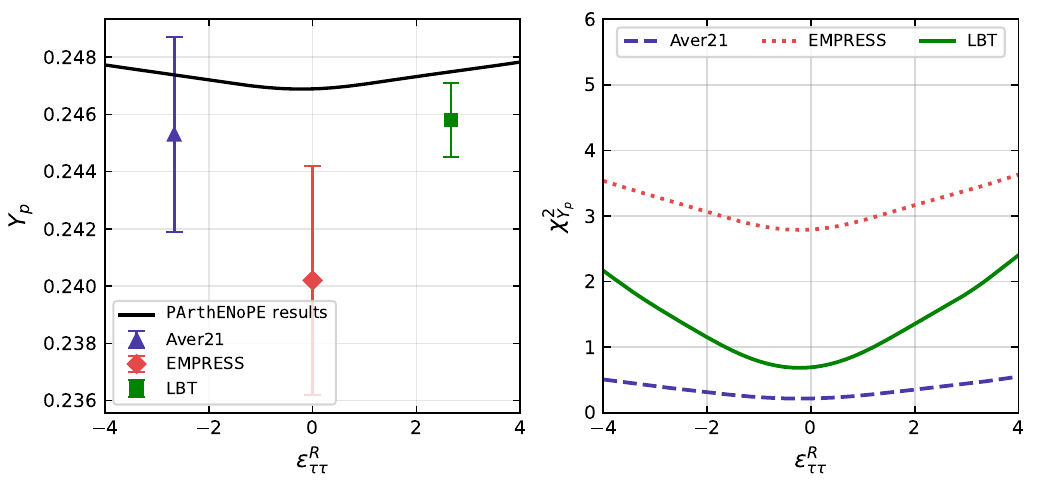}
    \caption{\emph{Left:} Primordial abundance of helium-4 as a function of $\varepsilon_{\tau \tau}^R$ (all other NSI parameters set to zero), compared to the observational determinations of Aver21~\cite{Aver:2020fon}, EMPRESS~\cite{Yanagisawa:2025mgx} and LBT~\cite{Aver:2026dxv} (note that they are displayed at different $\varepsilon_{\tau \tau}^R$ values solely for visualisation purposes). The black line is identical to the orange one on the bottom left panel of Fig.~\ref{fig:BBN_1D}. \emph{Right:} $\chi^2_{Y_p}$ [see Eq.~\eqref{eq:chi2}] when using each observational determination of $Y_p$.}
    \label{fig:Yp_chi2}
\end{figure}

In this Appendix, we focus on the role played in our analysis by the helium-4 measurement. In the main text, we have compared the 2021 determination by Aver \emph{et al.}~\cite{Aver:2020fon} (``Aver21'') and the more recent LBT result~\cite{Aver:2026dxv}. In recent years, the EMPRESS collaboration has reported \emph{smaller} values of $Y_p$~\cite{Matsumoto:2022tlr,Yanagisawa:2025mgx}, although the discrepancy has diminished in the latest determination,
\begin{equation}
\label{eq:Yp_EMPRESS}
    Y_p = 0.2402 \pm 0.0040 \, .
\end{equation}
In order to illustrate the impact of using these different measurements, we show in Fig.~\ref{fig:Yp_chi2}, for the one-parameter scan over $\varepsilon_{\tau \tau}^R$, how the $Y_p$ values compare to the observational bands (left panel), and the associated $\chi^2$ (right panel).

As already shown in Fig.~\ref{fig:PL_GR33}, the small range of variation of $Y_p$ makes it a limited probe of the NSI parameters, except when using the precise LBT measurement. The latest EMPRESS result~\eqref{eq:Yp_EMPRESS} has larger error bars than the Aver21 one (0.0040 vs 0.0034), yet one can see on the right panel of Fig.~\ref{fig:Yp_chi2} that it is better at constraining $\varepsilon_{\tau \tau}^R$. Instead of a precision-driven gain, it is a tension-driven one: the values of $\chi^2$ are significantly higher when using the EMPRESS likelihood, reflecting the distance between predictions and measurements on the left panel. Since these differences get less significant when adding the deuterium measurement to the likelihood, and even more the $\Neff$ constraint (in particular, as seen in Fig.~\ref{fig:PL_GR33}, only the LBT $Y_p$ value has better constraining power than the \emph{Planck} $\Neff$ constraint), we do not further elaborate on this point nor quote Bayesian credible intervals.

\bibliographystyle{JHEP}
\bibliography{bibliography.bib}

\end{document}